\documentclass{aa}
\usepackage{graphicx}
\usepackage{txfonts}
\usepackage{natbib}
\usepackage{color}
\bibpunct{(}{)}{;}{a}{}{,}    %to follow the A&A style 
\newcommand{\beq}{\begin{equation}}
\newcommand{\eeq}{\end{equation}}
\newcommand{\bea}{\begin{eqnarray}}
\newcommand{\eea}{\end{eqnarray}}
\newcommand{\lsim}{\raisebox{-0.6ex}{$\stackrel{{\displaystyle<}}{\sim}$}}

\newcommand{\url}[1]{{\tt #1}}

\def\gapp{\lower 3pt\hbox{${\buildrel > \over \sim}$}\ }
\def\lapp{\lower 3pt\hbox{${\buildrel < \over \sim}$}\ }

\def\AU{{\rm AU}}

\newlength{\linwx}
\begin{document}
\title{Orbital evolution of eccentric planets in radiative discs}
%  \subtitle{empty}
%%
\author{
Bertram Bitsch \inst{1}
and
Wilhelm Kley  \inst{1}
}
\offprints{B. Bitsch,\\ \email{bertram.bitsch@uni-tuebingen.de}}
\institute{
     Institut f\"ur Astronomie \& Astrophysik, 
     Universit\"at T\"ubingen,
     Auf der Morgenstelle 10, D-72076 T\"ubingen, Germany
}
\date{March 2010}
\abstract
%%  Context
{With an average eccentricity of about $0.29$, the eccentricity distribution of extrasolar planets is markedly
different from the solar system.
Among other scenarios considered, it has been proposed that eccentricity may grow through planet-disc interaction.
Recently, it has been noticed that the thermodynamical state of the disc can significantly influence the 
migration properties of growing protoplanets.
However, the evolution of planetary eccentricity in radiative discs has not been considered yet.
}
%%  Aims
{In this paper we study the evolution of planets on eccentric orbits that are embedded in a three-dimensional 
viscous disc and analyse the disc's effect on the orbital evolution of the planet.
}
%% Methods
{We use the three-dimensional hydrodynamical code
{{\tt NIRVANA}} that includes full tensor viscosity and 
implicit radiation transport in the flux-limited diffusion approximation.
The code uses the {\tt FARGO}-algorithm to speed up the simulations.
First we measure the torque and power exerted on the planet by the disc for fixed
orbits, and then we let the planet start with initial eccentricity and evolve it in the disc.
}
%% Results
{
For locally isothermal we confirm previous results and find eccentricity damping and inward migration
for planetary cores. For low eccentricity ($e \lsim 2 H/r$) the damping is exponential, while for higher $e$ it follows
$\dot{e} \propto e^{-2}$. In the case of radiative discs, the planets experience an inward migration as long as its eccentricity lies above a 
certain threshold. After the damping of eccentricity cores with masses below 33 $M_{Earth}$ begin to migrate outward in radiative discs, while
higher mass cores always migrate inward.
For all planetary masses studied (up to 200 $M_{Earth}$) we find eccentricity damping.
}
%% Conclusions
{
In viscous discs the orbital eccentricity of embedded planets is damped during the evolution independent of the mass.
Hence, planet-disc interaction does not seem to be a viable mechanism to explain the observed
high eccentricity of exoplanets.
}
\maketitle
\markboth
{Bitsch \& Kley: Orbital evolution of eccentric planets in radiative discs}
{Bitsch \& Kley: Orbital evolution of eccentric planets in radiative discs}

\section{Introduction}
\label{sec:introduction}

One of the surprising orbital characteristics of extrasolar planets is their high mean eccentricity {$\approx 0.29$}
\citep{2007ARA&A..45..397U}.
Several explanations have been put forward to explain this discrepancy in comparison to the solar system.
Planet-disc interactions are typically invoked to explain the planetary migration towards the central star
that has occurred during the early formation phase. In addition to the change in semi-major axis, it is to be expected
that the planet's eccentricity will be influenced through this process as well \citep{1980ApJ...241..425G}.
It has then been suggested, by performing linear analysis, that the planetary eccentricity can be increased through the planet-disc interaction under
some conditions \citep{2003ApJ...585.1024G,  2004ApJ...606L..77S, 2008Icar..193..475M}. 
They estimate that
eccentric Lindblad  resonances can  cause eccentricity  growth  for gap-forming planets.
Recently, a Kozai-type effect between the disc and an inclined
planet has been considered \citep{2010arXiv1001.0657T}.
Numerical simulations, however, tend to show predominantly
eccentricity damping for a variety of masses \citep{2007A&A...473..329C, 2009arXiv0904.3336M}.
Additionally, the existence of resonant planetary systems with relatively low eccentricities (such as the system GJ~876) seems
to indicate a damping action of the disc on planetary eccentricity rather than an enhancement
\citep{2002ApJ...567..596L,2005A&A...437..727K,2008A&A...483..325C}.

On the other hand, very high-mass planets can induce an eccentric instability in
the disc \citep{2006A&A...447..369K}. In turn, the eccentric disc can possibly increase the planetary eccentricity
\citep{2001A&A...366..263P,2006ApJ...652.1698D}. However, this process can only explain the eccentricity of
very massive ($\approx 5-10 M_{Jup}$) planets.
Alternatively, planet-planet scattering seems to be a viable mechanism for increasing eccentricities through mutual gravitational
interactions between the planets. The resulting eccentricity distribution matches the observed one
reasonably well \citep{2003Icar..163..290A, 2008ApJ...686..603J, 2008ApJ...686..621F}. Another option is the fly-by of a nearby
star \citep{2009MNRAS.394L..26M}.

Planet-disc interactions have so far been studied mostly in the locally isothermal approach, where the temperature only depends on
the distance from the central star. In this case for typical disc parameters, a negative torque is acting on the planet, and
it migrates inward  \citep{2002ApJ...565.1257T}. However, recently it has been shown that the torque acting on an embedded
planet depends on the thermodynamics of the disc. Following the pioneering work of \citet{2006A&A...459L..17P}, various groups 
have now analysed the effect of the equation of state on the migration properties \citep{2008ApJ...672.1054B, 2008A&A...485..877P, 2008A&A...478..245P, 2008A&A...487L...9K}.
Through full 3D radiative simulations of embedded planets, we have recently confirmed
that including of radiation transport can produce a positive torque acting on low-mass planets embedded in protoplanetary discs
\citep{2009A&A...506..971K}, 
because through its action the required radial entropy gradient can be maintained in the disc. 
This results in slowing down the inward migration, and 
under some conditions it may indeed be possible to reverse the inward migration process. 

The linear estimates of the eccentricity evolution of embedded planets
\citep{1993ApJ...419..166A, 1994Icar..110...95W, 2004ApJ...602..388T} concentrate
on low eccentricities and predict exponential decay on short timescales
$\tau_{ecc} \approx (H/r)^2 \tau_{mig}$, where $H/r$ is the aspect ratio of the disc
and $\tau_{mig}$ and $\tau_{ecc}$ the migration and eccentricity damping timescale, respectively.
\citet{2000MNRAS.315..823P} have also considered larger values for $e$ and they find an extended eccentricity
damping timescale such that $de/dt \propto e^{-2}$ if $e > 1.1 H/r$. 
\citet{2006A&A...450..833C} have performed hydrodynamical
simulations of embedded small mass planets and find good agreement with the work
by \citet{2000MNRAS.315..823P}.
These 2D results have been confirmed by \citet{2007A&A...473..329C} using fully 3D isothermal simulations.

As mentioned above, the thermodynamics of protoplanetary discs is a crucial parameter for the torque acting on the planet \citep{2006A&A...459L..17P}.
Including radiation transport/cooling in a disc will give rise to positive torques acting on a planet embedded in such a disc,
which indicates outward migration. In our previous work \citep{2009A&A...506..971K} we have shown that the inclusion of
radiation transport/cooling in simulations with embedded low mass planets in three-dimensions (3D) can result in outward migration.
So far these simulations have been limited to planets on fixed circular orbits.
Now we extend this work and focus on the evolution of planets on eccentric orbits for both, the isothermal and fully radiative regime.
We focus first on low-mass planets and study
the influence of the thermodynamics of the disc on eccentricity damping as well as on the evolution of the planet inside the disc.
First we estimate the theoretical migration and eccentricity damping rate for planets on fixed eccentric orbits.
Secondly, we let the planets evolve in the disc and finally we investigate the influence of the planet mass on the
change in eccentricity and semi-major axis.

\section{Physical modelling}
\label{sec:model}

The protoplanetary disc is modelled as a three-dimensional (3D), non-self-gravitating gas whose motion is described by the Navier-Stokes equations. We treat the disc as a viscous medium, where the dissipative effects can then be described via the standard viscous stress-tensor approach \citep[eg.][]{1984frh..book.....M}. We also assume that the heating of the disc occurs solely through internal viscous dissipation and ignore the influence of additional energy sources (e.g. irradiation form the central star). This internally produced energy is then radiatively diffused through the disc and eventually emitted from its surface. For this process we use the flux-limited diffusion approximation \citep[FLD,][]{1981ApJ...248..321L}, which allows us to treat the transition form optically thick to thin regions approximately. The viscous forces used in our code are stated explicitly for the three-dimensional case in spherical coordinates in \citet{1978trs..book.....T}. We use a constant kinematic viscosity coefficient $\nu$
with a dimensionless value of $\nu=10^{-5}$ (in code units, see below). 
This relates to the typically used $\alpha$-parameter through
$\nu = \alpha c_s H$, where $c_s$ is the sound speed and $H$ the vertical thickness of the disc.
A more  detailed prescription of the modelling and the 
numerical methodology is described in our previous paper \citep{2009A&A...506..971K}.
We now extend the simulations, compared to our previous paper, by including planets on eccentric orbits with different masses.
%% from a one-sided disc (symmetric with respect to the midplane) to a two-sided disc for our simulations with inclined orbits.

\subsection{General Setup}
An important issue in modelling planetary dynamics in discs is the gravitational potential of the planet since this has
to be artificially smoothed to avoid singularities. 
We have shown in \citep{2009A&A...506..971K} that the physics of embedded planets
can be described better by a cubic-potential rather than the often used $\epsilon$-potential.
Hence, we use in this work the following form for the planetary potential throughout
\begin{equation}
\label{eq:cubic}
\Phi_p^{cub} =  \left\{
    \begin{array}{cc} 
   - \frac{m_p G}{d} \,  \left[ \left(\frac{d}{r_\mathrm{sm}}\right)^4
     - 2 \left(\frac{d}{r_\mathrm{sm}}\right)^3 
     + 2 \frac{d}{r_\mathrm{sm}}  \right]
     \quad &  \mbox{for} \quad  d \leq r_\mathrm{sm}  \\
   -  \frac{m_p G}{d}  \quad & \mbox{for} \quad  d > r_\mathrm{sm} 
    \end{array}
    \right.
\end{equation}
Here $m_P$ is the planetary mass, $d=| \mathbf{r} - \mathbf{r_P}|$ denotes the distance of the disc element to the planet and $r_{sm}$ is the smoothing length of the potential measured in units of the Hill radius. The construction of the planetary potential is in such a way that for distances larger than $r_{sm}$ the potential matches the correct $1/r$ potential and is smoothed inside that radius ($d < r_{sm}$) by a cubic polynomial. The parameter $r_{sm}$ is equal to $0.5$ in all our simulations, unless stated otherwise.

The gravitational torques acting on the planet are calculated by integrating over the whole disc, where we apply a tapering function to exclude the inner parts of the Hill sphere of the planet. Specifically, we use the smooth (Fermi-type) function
\begin{equation}
\label{eq:fermi}
     f_b (d)=\left[\exp\left(-\frac{d/R_H-b}{b/10}\right)+1\right]^{-1}
\end{equation}
which increases from 0 at the planet location ($d=0$) to 1 outside $d \geq R_{H}$ with a midpoint $f_b = 1/2$ at $d = b R_{H}$, i.e. the quantity $b$ denotes the torque-cutoff radius in units of the Hill radius. This torque-cutoff is necessary to avoid large, probably noisy contributions from the inner parts of the Roche lobe and to disregard material that is gravitational bound to the planet \citep{2009A&A...502..679C}.
Here we assume (as in our previous paper) a transition radius of $b= 0.8$ Hill radii.

\subsection{Initial Setup}

The three-dimensional ($r, \theta, \phi$) computational domain consists of a complete annulus of the protoplanetary disc centred on the star, extending from $r_{min}=0.4$ to $r_{max}=2.5$ in units of $r_0=a_{Jup}=5.2 AU$. In the vertical direction the annulus extends from the disc's midplane (at $\theta = 90^\circ$) to $7^\circ$ (or $\theta = 83^\circ$) above the midplane for the simulations of planets with eccentric orbits in the midplane of the disc.
Here $\theta$ denotes the polar angle of our spherical polar coordinate system measured from the polar axis.
The central star has one solar mass $M_\ast = M_\odot$, and the total disc mass inside [$r_{min}, r_{max}$] is $M_{disc} = 0.01 M_\odot$. For the isothermal simulations we assume an aspect ratio of $H/r=0.037$ for the disc, in very close agreement with the
fully radiative models of our previous studies. For the radiative models $H/r$ is calculated self-consistently from the equilibrium structure
given by the viscous internal heating and radiative diffusion.
The isothermal models are initialised with constant temperatures on cylinders with a profile $T(s) \propto s^{-1}$ with $s=r \sin \theta$. This yields a constant ratio of the disc's vertical height $H$ to the radius $s$. The initial vertical density stratification is approximately given by a Gaussian:
\begin{equation}
  \rho(r,\theta)= \rho_0 (r) \, \exp \left[ - \frac{(\pi/2 - \theta)^2 \, r^2}{2 H^2} \right] \ .
\end{equation} 
Here, the density in the midplane is $\rho_0 (r) \propto r^{-1.5}$ which leads to a $\Sigma(r) \propto \, r^{-1/2}$ profile of the vertically integrated surface density. In the radial and $\theta$-direction we set the initial velocities to zero, while for the azimuthal component the initial velocity $u_\phi$ is given by the equilibrium of gravity, centrifugal acceleration and the radial pressure gradient. This corresponds to the equilibrium configuration for a purely isothermal disc. 

For our fully radiative model we first run a 2D axisymmetric model (starting from the given isothermal equilibrium) to obtain a new self-consistent equilibrium where viscous heating balances radiative transport/cooling from the surfaces. After reaching that equilibrium, we extend this model to a full 3D simulation, by expanding the grid into $\phi$-direction. The resulting disc for this model has $H/r \approx 0.037$ so we choose that value
for our isothermal runs.

\subsection{Numerical Setup}

Our coordinate system rotates at the initial orbital frequency of the planet (at $r=r_0$). We use an equidistant grid in $r,\theta,\phi$ with a resolution of ($N_r,N_\theta,N_\phi)=(266,32,768)$ active cells for our simulations. At $r_{min}$ and $r_{max}$ we use damping boundary conditions for all three velocity components to minimise disturbances (wave reflections) at these boundaries. The velocities are relaxed towards their initial state on a timescale of approximately the local orbital period. The angular velocity is relaxed towards the Keplerian values, while the radial velocities at the inner and outer boundaries vanish. Reflecting boundary conditions are applied for the density and temperature in the radial directions. We apply periodic boundary conditions for all variables in the azimuthal direction. In the vertical direction we set outflow boundary conditions for $\theta_{min}$ (the surface of the disc) and symmetric boundary conditions at the disc's midplane 
($\theta_{max} = \pi/2$). 
We use the finite volume code {\tt NIRVANA} \citep{1997ZiegYork} with implicit radiative transport in the flux-limited
diffusion approximation and the {\tt FARGO} extension as described in \citet{2009A&A...506..971K}.

\subsection{Simulation Setup}

In the first part of our model sequence we consider the orbital evolution of a planet with a fixed mass ($20 M_{Earth}$) on
eccentric orbits using different initial eccentricities. 
For comparison we consider isothermal and fully radiative models. 
Using radiative discs here is a direct extension of a previous study under purely isothermal disc conditions using the same
planet mass \citep{2007A&A...473..329C}.
We distinguish two different approaches for these $20 M_{Earth}$ models: 
first, a model sequence where the planet stays on a fixed eccentric orbit and secondly where the planet is free to move inside the computational domain under the action of the planet-disc gravitational forces. 
For the second models we insert the planet in the disc and let it move immediately, but using a time-dependent mass growth of the planet (through the planetary potential) until it reaches its destination mass. 
For the first set of models the $20 M_{Earth}$ planet is inserted as a whole in the disc at the start of the simulation.
Initially the planet starts at a distance $r=a_{Jup}=5.2 AU$ from the central star. 
For the fully radiative simulations we set the ambient temperature to a fixed value of $10 K$ at the disc surface
(at $\theta_{min}$), which ensures that all the internally generated energy is liberated freely at the disc's surface.
This low temperature boundary condition works very well at optically thin boundaries and does not effect the inner parts of the optically thick disc \citep{1999ApJ...518..833K, 2009A&A...506..971K}.
In the second part of the project we consider sequences of models for a variety of planet masses.
We note, that a $20 M_{Earth}$ planet has in our simulations using our standard resolution a Roche radius of about $3.3$ grid cells.
As we will see later in the results section, there is indication that for small mass planets and isothermal runs
(using the cubic-potential) a higher resolution is required.

\section{Models with an embedded planet on fixed eccentric orbits}
\label{sec:eccentric}

In this section we consider planets remaining on fixed eccentric orbits embedded in either isothermal or fully radiative discs.
From the disc forces acting on the planet we calculate its theoretical migration rate and eccentricity change.
Below we will compare this directly to moving planets in the isothermal and fully radiative regime.

\subsection{Torque and Power}

From the gravitational forces acting on the planet we can calculate the torque and energy loss (power) of the planet. These can be used
to estimate the theoretical change of the eccentricity and the semi-major axis of the planet.
Here we follow \citet{2007A&A...473..329C}.

The angular momentum $L_p$ of a planet on an eccentric orbit is given by
\begin{equation}
	L_p = m_p \sqrt{G M_\ast a} \sqrt{1-e^2} \ .
\end{equation}
In our particular case, for non-inclined planets $L_p = L_z$. We can then obtain the rate of change of the semi-major axis and the eccentricity by
\begin{equation}
\label{eq:semiecc}
	\frac{\dot{L}_p}{L_p} = \frac{1}{2} \frac{\dot{a}}{a} - \frac{e^2}{1-e^2} \frac{\dot{e}}{e} = \frac{T_{disc}}{L_p} \ ,
\end{equation}
where $T_{disc}$ is the total torque exerted by the disc onto the planet
\begin{equation}
 	T_{disc} = \int_{disc} (\mathbf{r}_p \times \mathbf{F} ) \ |_z \, dV \ .
\end{equation}
Here $\mathbf{r}_p$ denotes the radius vector from the star to the planet, $\mathbf{F}$ the (gravitational) force per unit volume between the planet and a disc element (at location $\mathbf{r}$ from the star), and $dV$ the volume element. Equation (\ref{eq:semiecc}) implies that a positive torque may also result in eccentricity damping rather than outward migration. This is actually the observed result for our moving planets.

The energy change per time (power) of the planet due to the work done by the gravitational forces of the disc is given by
\begin{equation}
	P_{disc} = \int_{disc} \mathbf{r}_p \cdot \mathbf{F} \ dV \ .
\end{equation}
The energy of the planet depends only on the semi-major axis $a$ of the planet and is given by
\begin{equation}
 	E_p = - \frac{1}{2} \frac{G M_\ast m_p}{a} \ .
\end{equation}
We can now obtain for the energy loss and semi-major axis change
\begin{equation}
\label{eq:aeng}
	\frac{\dot{E}_p}{|E_p|} = \frac{\dot{a}}{a} = \frac{P_{disc}}{|E_p|} \ .
\end{equation}
From equation (\ref{eq:aeng}) and equation (\ref{eq:semiecc}) we can calculate the theoretical change of the semi-major axis and the eccentricity for planets on a fixed eccentric orbit for the isothermal and fully radiative cases. We find for the theoretical change of the semi-major axis
\begin{equation}
\label{eq:dota}
 \frac{\dot{a}}{a} = \frac{P_{disc}}{|E_P|} = \frac{2 a}{G M_\ast m_p} \, P_{disc} 
\end{equation}
and for the change of eccentricity
\begin{equation}
\label{eq:dote}
 \frac{\dot{e}}{e} = \frac{1-e^2}{e^2} \left( \frac{1}{2} \frac{\dot{a}}{a} - \frac{T_{disc}}{L_P} \right)
\end{equation}
with our calculated $\dot{a}$.

\begin{figure}
 \centering
 \includegraphics[width=0.9\linwx]{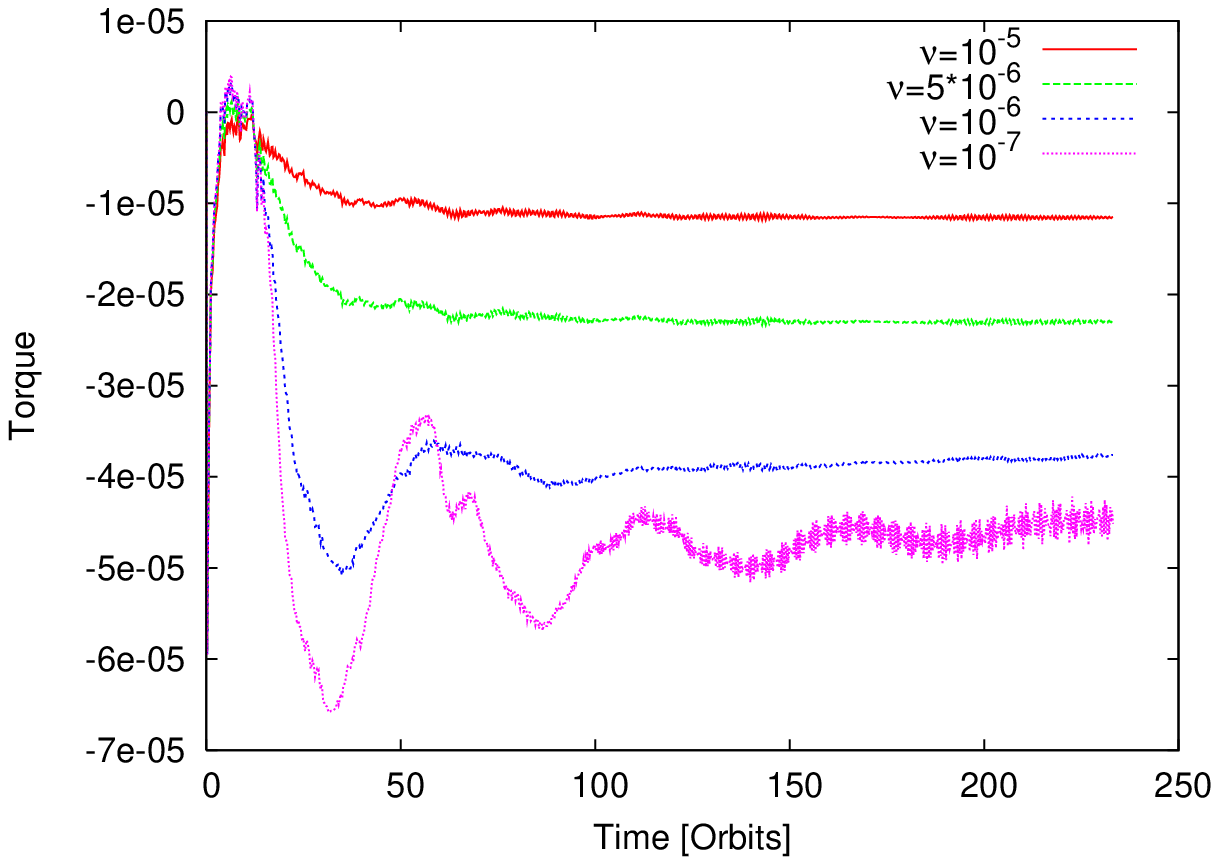}
 \includegraphics[width=0.9\linwx]{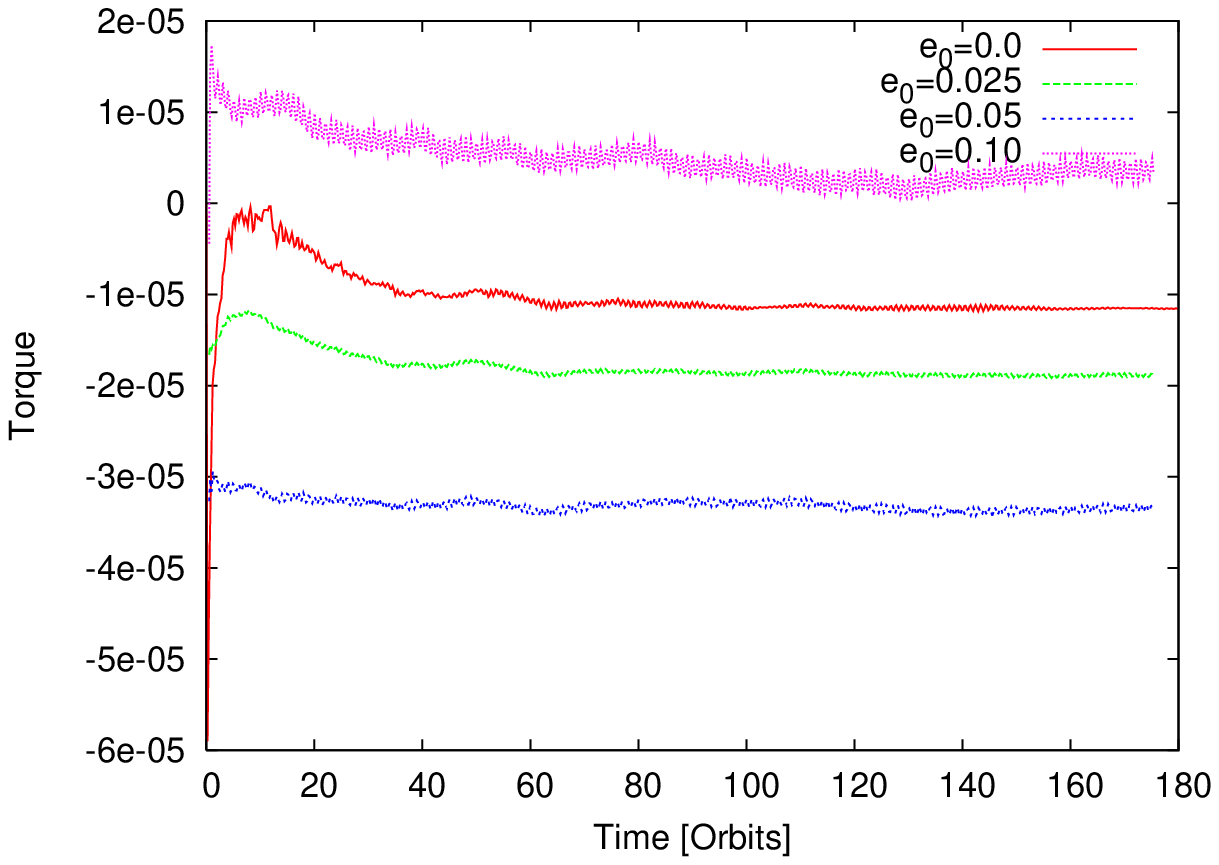}
 \caption{Time dependence of the torque for different isothermal models with planets on fixed orbits. Top: 
  Circular planets for different values of the viscosity. The curves are ordered from high to small viscosity.
  For smaller viscosities the settling towards equilibrium takes longer. Bottom:
  Evolution for our standard viscosity ($\nu = 10^{-5}$) for different eccentricities.
  Due to the large time variability of the torque for eccentric orbits these curves use sliding time averaged values with
  a window width of 1 period. 
   \label{fig:Torqtime}
   }
\end{figure}

\begin{figure}
 \centering
 \includegraphics[width=0.9\linwx]{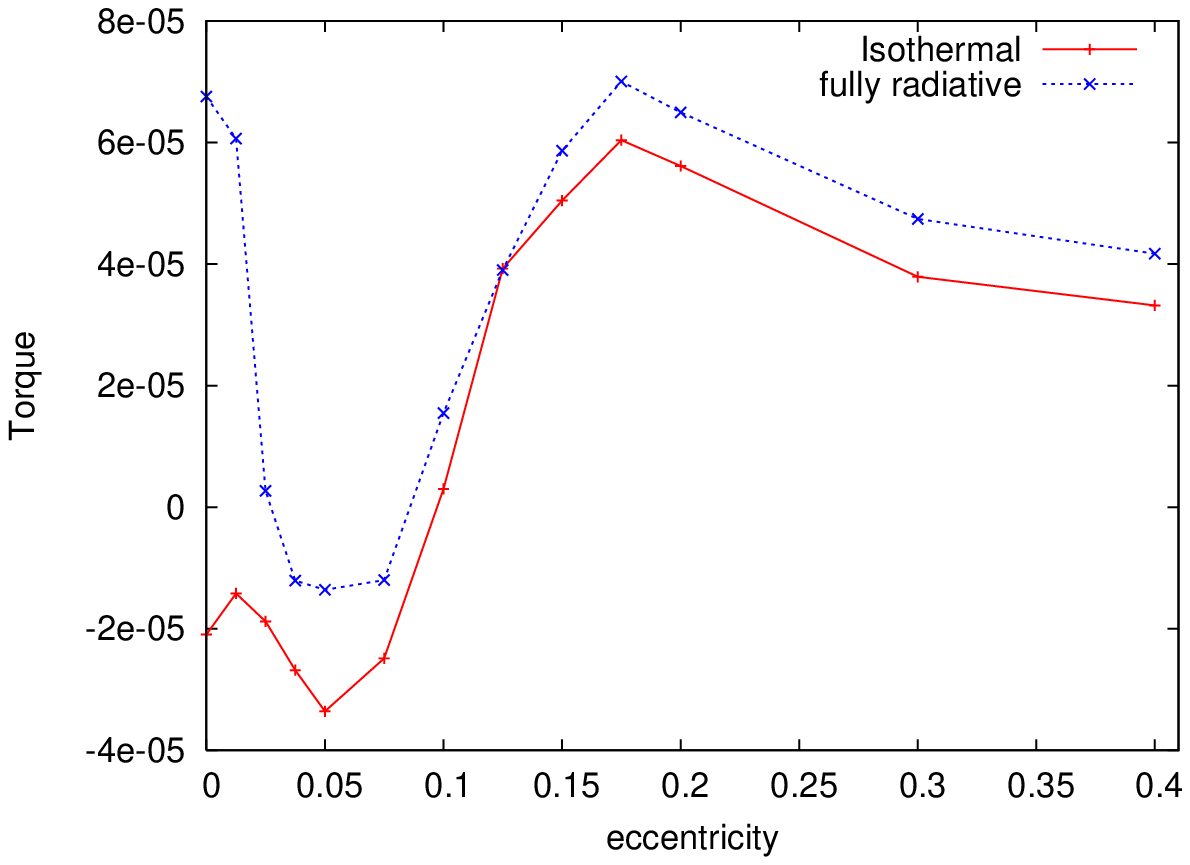}
 \includegraphics[width=0.9\linwx]{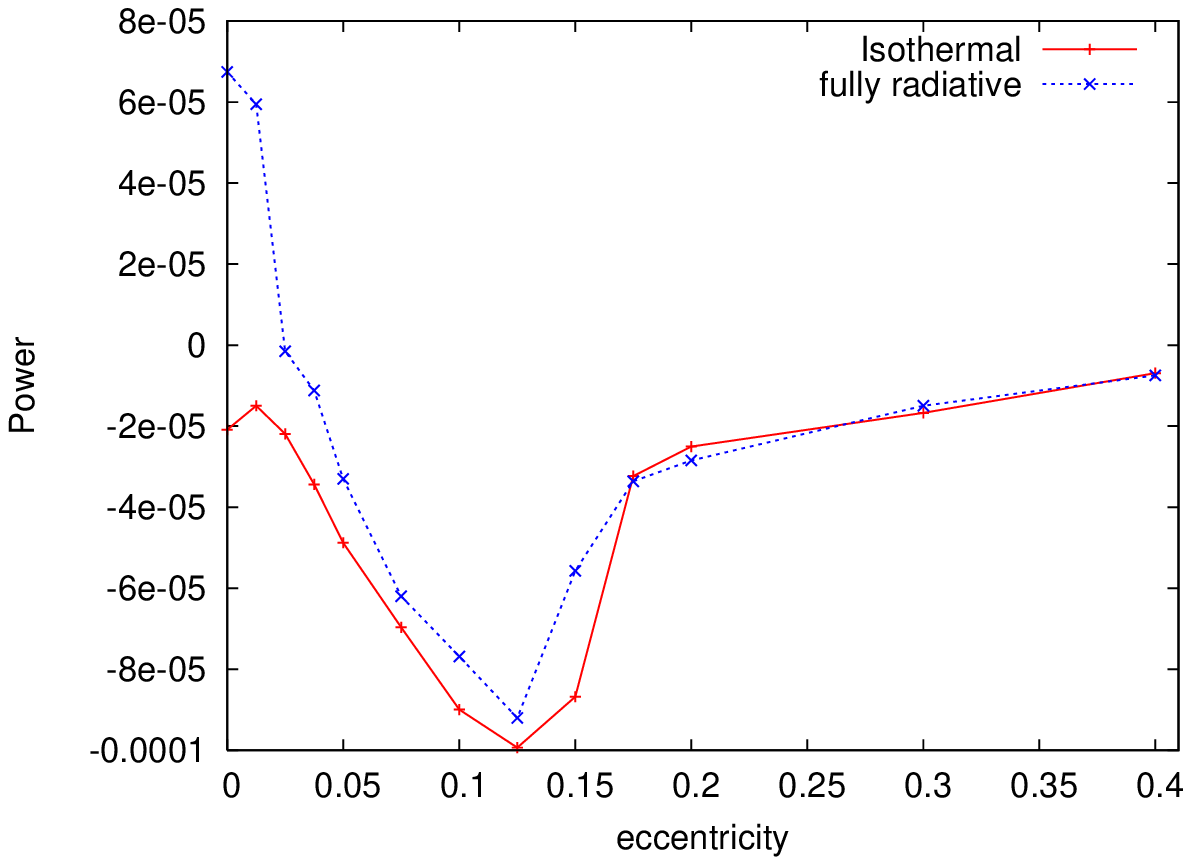}
 \caption{Torque (top) and power (bottom) acting on a planet on a fixed eccentric orbit in dependency of the eccentricity and thermodynamics of the planet. The torque and the power have been averaged over $20$ orbits, taken from $t=140$ to $t=160$ orbits for the isothermal (solid red line) and from $t=120$ to $t=140$ for the fully radiative simulation (dashed blue line). 
   \label{fig:EcctorIsofull}
   }
\end{figure}

Our simulations with planets on a fixed eccentric orbit feature eccentricities ranging from $e_0=0.0125$ to $e_0=0.4$.
To investigate a possible change in the orbital elements of a planet we first analyse the torques and power acting on the planet
for fixed eccentric orbits.
The time evolution of the torques for different models is displayed in Fig.~\ref{fig:Torqtime}. The top panel refers to
planets on circular orbits for different viscosities, as quoted in the caption. For the smallest viscosity ($\nu = 10^{-7}$) the torque
is unsaturated and evolves through long period oscillations towards the equilibrium value. The timescale of the oscillation
is comparable to the libration time of a particle near the edge of the horsehoe region (see App.~\ref{app:Numfeature}).
For larger viscosity the equilibration time becomes shorter as the
the viscous diffusion time shortens. The results are in very good agreement with existing 2D simulations 
\citep{2009MNRAS.394.2283P,2010MNRAS.401.1950P}, and confirm
clearly that viscosity is a necessary ingredient for torque saturation.
Additionally, this result indicates that the intrinsic (numerical) diffusivity is much smaller than that given by our standard physical viscosity.
In the lower panel of Fig.~\ref{fig:Torqtime} we display similar curves for eccentric planets and the standard viscosity
$\nu = 10^{-5}$. Due to the strong variability of the torque on the orbital timescale for eccentric orbits (see \citet{2007A&A...473..329C} and
below) we display time averaged torques.

In Fig.~\ref{fig:EcctorIsofull} we display the torque and power acting on such a planet for isothermal and fully radiative runs in dependency of the eccentricity of the planet. 
As only every $10$th time step was plotted in our output file we averaged the torque over $20$ orbits to minimise the numerical fluctuations due to this procedure.
For eccentricities smaller than $0.1$ the planet experiences a negative torque in the isothermal simulations, while for higher eccentricities the planet feels a positive torque. The torque reaches a maximum at $e=0.175$ and settles down to a nearly constant value for larger eccentricities, which is in good agreement with \citet{2007A&A...473..329C}. The differences in the absolute values of their torque compared to ours have their origin in the aspect ratio $H/r$ of the disc.
Our torque is generally higher, as a result of our lower $H/r=0.037$ compared to their $H/r=0.05$ disc. 

The torque acting on the planet is in general higher for the fully radiative simulations than for the isothermal ones. This phenomenon was observed in various simulations in the past for planets staying on fixed circular orbits, so it is no surprise that the torque for the fully radiative simulation is higher for planets on eccentric orbits as well. 
For low eccentric planets the fully radiative simulation yield positive torques (and power) in agreement with our previous
results for planets on purely circular orbits.
On circular orbits, inclusion of radiation transport/cooling, gives rise to a positive torque, which implies outward migration in contrast to the isothermal simulations. The effect is caused by corotation region that gives rise to a positive contribution to the torque in the case of a positive entropy gradient. 
Interesting is the narrow range in eccentricity
of this outward migration. Already for $e \approx 0.03$ the direction of migration is directed inward again. This effect is caused by the
spatially narrow region that contributes to the positive torque on the planet \citep{2009A&A...506..971K}.
For larger eccentricities the isothermal and radiative results match reasonably well, but the radiative results are always
slightly larger than the isothermal ones. Different effects can contribute to this offset: the difference in sound speed, the spatially varying
$H/r$ for the radiative runs in contrast to the constant value for the isothermal runs, or a difference in the corotation  torques.

In the bottom diagram of Fig.~\ref{fig:EcctorIsofull} the displayed power of the planet follows the trend of the diagram of the torque acting on the planet with one big difference: the power of the planet is always negative for the isothermal simulations (implying inward migration), while it is positive for low eccentric planets in the fully radiative scheme. For small $e$ the torque and power are very similar for both cases, since they must be identical for $e=0$.

\begin{figure}
 \centering
 \includegraphics[width=0.9\linwx]{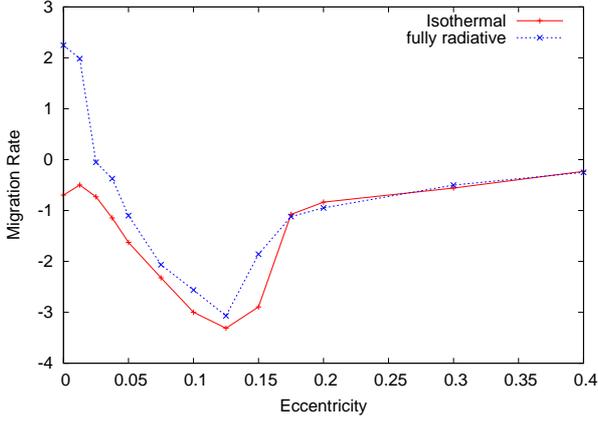}
 \caption{Theoretical rate of change of the semi-major axis ($\dot{a}/a$) for eccentric planets on fixed orbits. For eccentricities smaller than about $0.025$ we find a positive migration rate for the fully radiative disc, which indicates outward migration. This can actually been observed for moving planets later in the text.
   \label{fig:AIsofullrate}
   }
\end{figure}

\begin{figure}
 \centering
 \includegraphics[width=0.9\linwx]{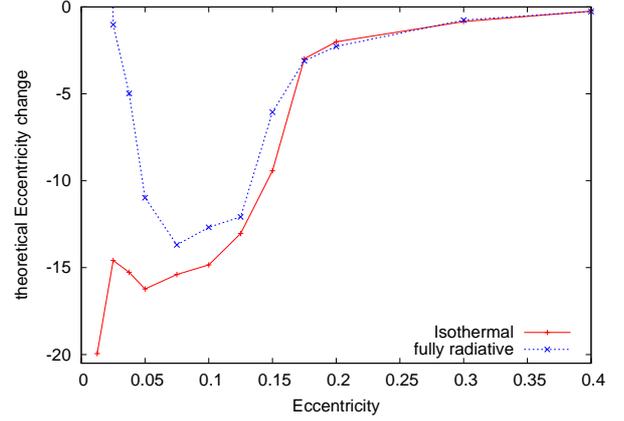}
 \caption{Theoretical rate of change of the eccentricity ($\dot{e}/e$) for eccentric planets on fixed orbits.
   \label{fig:EccIsofullrate}
   }
\end{figure}

The consequences for the inferred change in semi-major axis and eccentricity are displayed in Figs.~\ref{fig:AIsofullrate} and \ref{fig:EccIsofullrate}.
The theoretical migration rate (Eq.~\ref{eq:dota}) for planets on fixed eccentric orbits (Fig.~\ref{fig:AIsofullrate}) reflects our assumptions. When a planet has a high initial eccentricity ($e>0.2$) the migration rate is nearly constant and inward for the isothermal and fully radiative case, but as soon as the eccentricity gets damped to a value smaller than $e=0.2$ the inward migration increases by a factor of $2$ to $3$. The fastest inward migration is seen for planets with an eccentricity of $e \approx 0.125$ for both thermodynamic cases. If the eccentricity evolves to lower values this rapid inward migration is slowed down in the isothermal case. In the fully radiative case this process is even stronger, so that planets with a very low eccentricity ($e \leq 0.025$) have a positive migration rate, indicating outward migration. The positive migration rate is a consequence of the positive torque and power acting on the planet. This confirms very well our previous work of low-mass planets on circular orbits in fully radiative discs migrating outward \citep{2009A&A...506..971K}. The causes for this outward migration are the same as for circular orbits, as we will see later on. The migration rate in the isothermal case is faster for low eccentric planets (with $e\approx 0.125$) and a little bit slower for high eccentric planets compared to the zero eccentricity case.

%In the isothermal case, the migration rate is negative for all eccentricities with a maximum loss at $e=0.10$ which can be observed in the evolution of the semi-major axis (Fig.~\ref{fig:AIsoEccall}). In the beginning the simulations with planets starting at $e=0.10$ and $e=0.15$ have the greatest loss of semi-major axis, which changes in time, as the eccentricity of the moving planet changes. For planets with high initial eccentricity the migration rate is somewhat slower that for these low eccentric planets. Is the eccentricity of an initial high eccentric planet damped to about $e\approx 0.15$ the migration rate becomes faster (see Fig.~\ref{fig:AIsoEccall}) according to our theoretical migration rate, but as the eccentricity is damped even more the migration rate becomes slower again.

%In the fully radiative regime planets with $e < 0.025$ have a positive migration rate right from the start, indicating outward migration or gain of eccentricity. In in Fig.~\ref{fig:AfullEccall} we can clearly see that the planet moves outward in the fully radiative scheme. For bigger eccentricities the migration rate follows the isothermal migration rate, meaning that planets with higher initial eccentricity first move inward, as the eccentricity is damped. The damping procedure for high eccentric planets is in principal the same as for the isothermal regime. As soon as the planets eccentricity is damped to $e < 0.02$ the migration rate changes its sign, and becomes positive, pushing the planet outward as we have seen in our simulations.

The theoretical damping of eccentricity in the isothermal case (Fig.~\ref{fig:EccIsofullrate}) indicates eccentricity damping for all
values of $e$, in agreement with our results for moving planets (see Fig.~\ref{fig:EccIsoEccall} below). For low eccentricity planets ($e \leq 0.10$) the damping of eccentricity is much faster than for high eccentric planets. As soon as the eccentricity of high eccentric planets is damped to a low eccentricity, the damping of eccentricity becomes faster, but will then come to a constant value. In the fully radiative case, the damping of eccentricity is slower for low eccentric planets ($e \leq 0.10$) compared to the isothermal simulations but is nearly the same for high eccentric planets. This means the damping rate for eccentric planets is somewhat bigger in the fully radiative case compared to the isothermal case. For very small eccentricities we even get a positive value for the change of eccentricity (cut off in the Figure) in the fully radiative case. In the isothermal case we note the opposite effect: for very small eccentricities the change of eccentricity is somewhat larger than for slightly higher eccentricities. This phenomenon might have its origin in a numeric feature: for small eccentricities the calculations of the migration rate and the change of eccentricity become very sensitive. Interestingly the slower change in eccentricity for the fully radiative simulations results in a lower final eccentricity.
For eccentricities smaller than about $0.025$ we find a positive eccentricity change in the fully radiative case, which is actually not seen in our simulations of moving planets, and is a result of small numerical inaccuracy combined with the division by $e$ in Eq.~\ref{eq:dote}.

\subsection{Torque analysis}

\begin{figure}
 \centering
 \includegraphics[width=0.9\linwx]{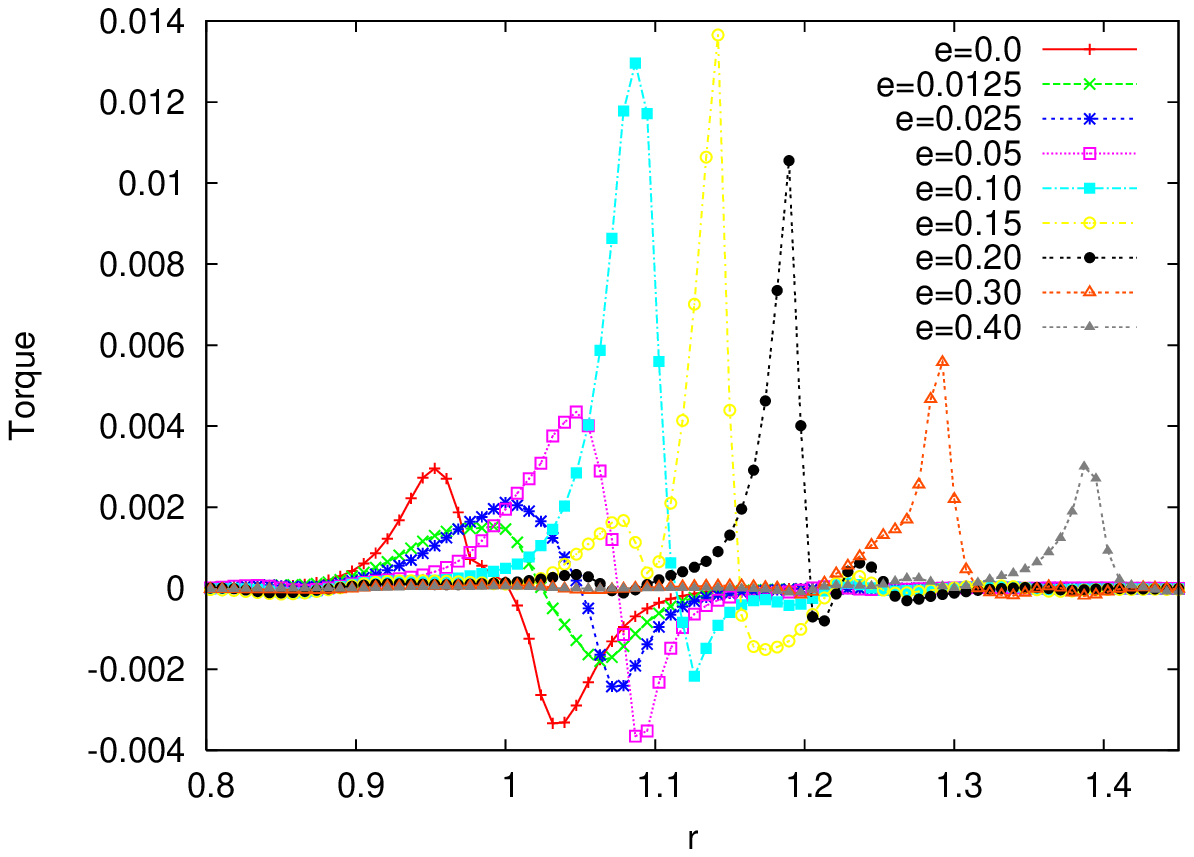}
 \includegraphics[width=0.9\linwx]{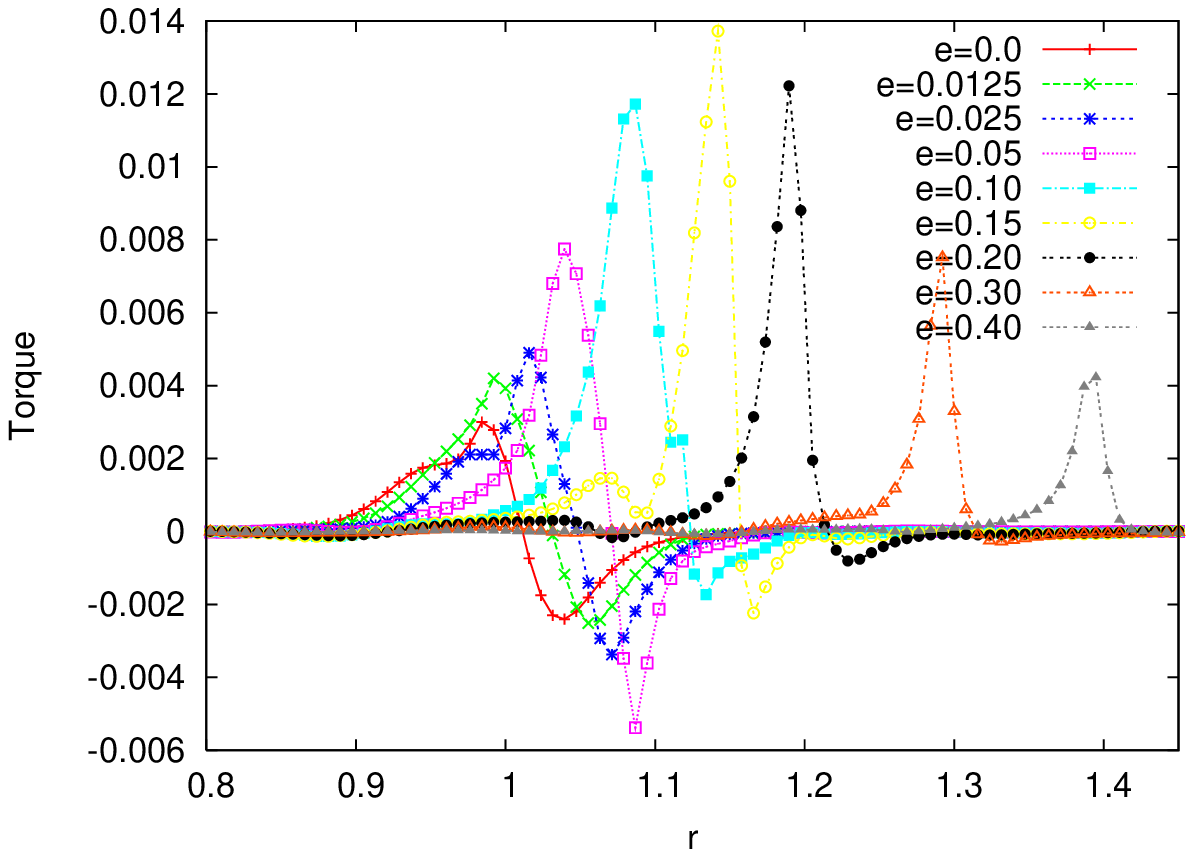}
 \caption{Torque density acting on a planet on a fixed eccentric orbit in dependency of the eccentricity of the planet embedded in an isothermal disc (top) and a fully radiative disc (bottom). The snapshots are taken at $t=150$ Orbits for the isothermal simulations and at $t=100$ Orbits for the fully radiative simulations. The planet is located at apoastron in all cases. 
   \label{fig:EccGammaIsofull}
   }
\end{figure}

To understand the behaviour of the total torque in more detail we analyse now the space-time variation of the torque and power of the planet.
For that purpose we introduce the radial torque density $\Gamma (r)$, which is defined in such a way that the total torque $T^{tot}$ acting on the planet is given by
\beq
       T^{tot} = \int_{r_{min}}^{r_{max}} \, \Gamma(r) \, dr \ .
\eeq
The radial torque density has been a very useful tool to investigate the origin of the torques in our previous work on planets on fixed circular orbits. In Fig.~\ref{fig:EccGammaIsofull} we display $\Gamma (r)$ for a selection of our planets. All the snapshots were taken when the planet
was located at apoastron. Note, that $\Gamma(r)$ changes during the orbit, as the planet moves on an eccentric orbit.

\begin{figure}
 \centering
 \includegraphics[width=0.9\linwx]{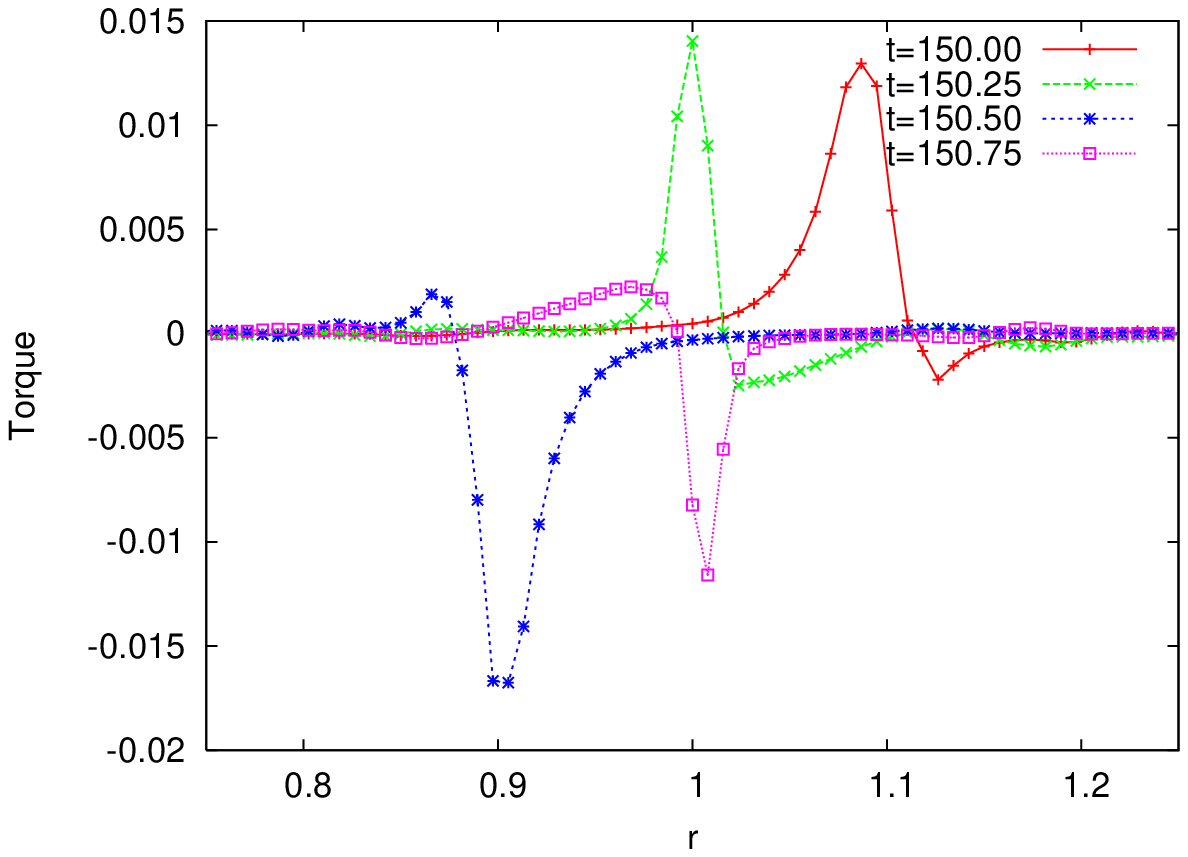}
 \includegraphics[width=0.9\linwx]{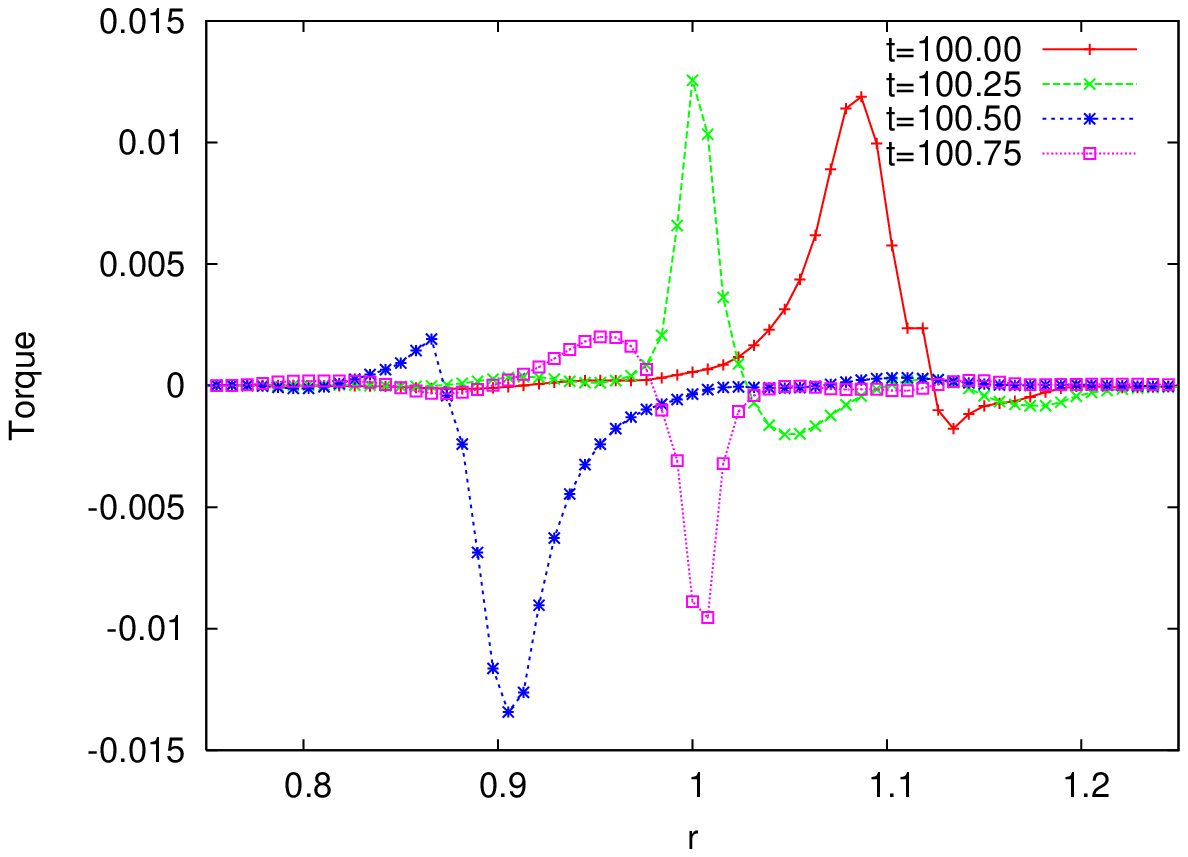}
 \caption{Torque density acting on a planet on a fixed eccentric orbit ($e=0.10$) during the time of one planetary orbit in an isothermal disc (top) and in a fully radiative disc (bottom). One can clearly see that the torque changes according to the position of the planet to the central star, making it clear why the torques have to be averaged.
   \label{fig:EccGammatimeIsofull}
   }
\end{figure}

On can clearly see, that the major contribution to the torque originates at larger radii for larger eccentricities by construction, as the planets location is at apoastron and its distance from the central star is $r=a+e$, with $a=1.0$. For eccentricities smaller than $e=0.1$ Fig.~\ref{fig:EccGammaIsofull} suggests a negative total torque, while for larger eccentricities a positive torque is assumed, which can be clearly seen in Fig.~\ref{fig:EcctorIsofull}. One might also argue, that the torque acting on the planet for eccentricities larger than $e=0.1$ should be much higher than shown in Fig.~\ref{fig:EcctorIsofull}. But keep in mind that the torque acting on the planet changes during one orbit as the planet moves on eccentric orbits, see Fig.~\ref{fig:EccGammatimeIsofull}.

In Fig.~\ref{fig:EccGammatimeIsofull} the planet has initially a distance of $r=1.1$ to the central star (the planet is located at apoastron) and after half an orbit it is nearest to the central star (at $r=0.9$, the planet is located at periastron) and moves then further away from the central star to $r=1.1$ again. The motion of the planet in respect to the central star is the reason for the change in the  $\Gamma(r)$-function with respect to time.

\begin{figure}
 \centering
 \includegraphics[width=0.823\linwx]{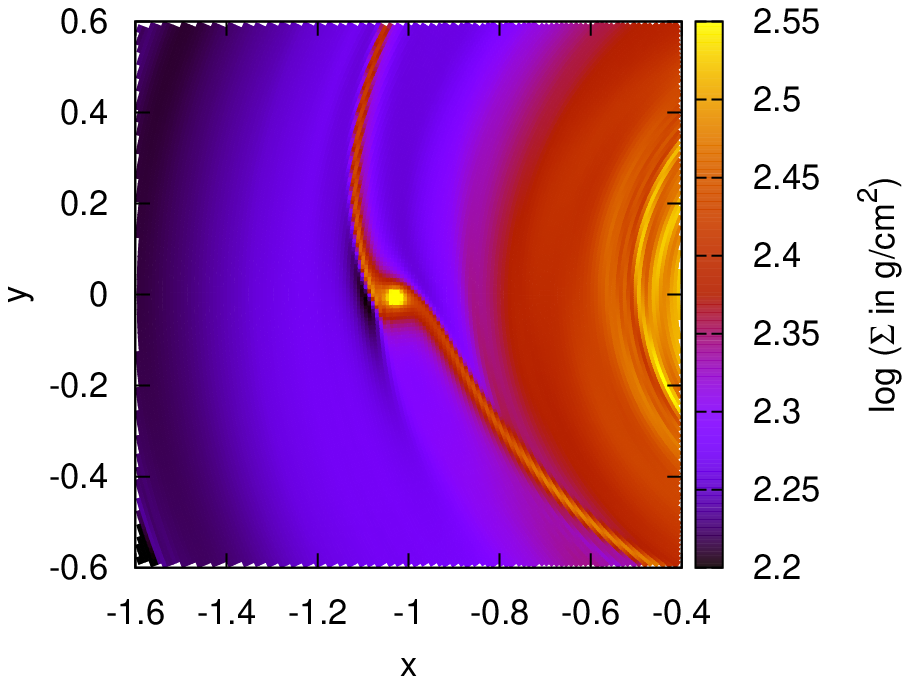}

 \includegraphics[width=0.823\linwx]{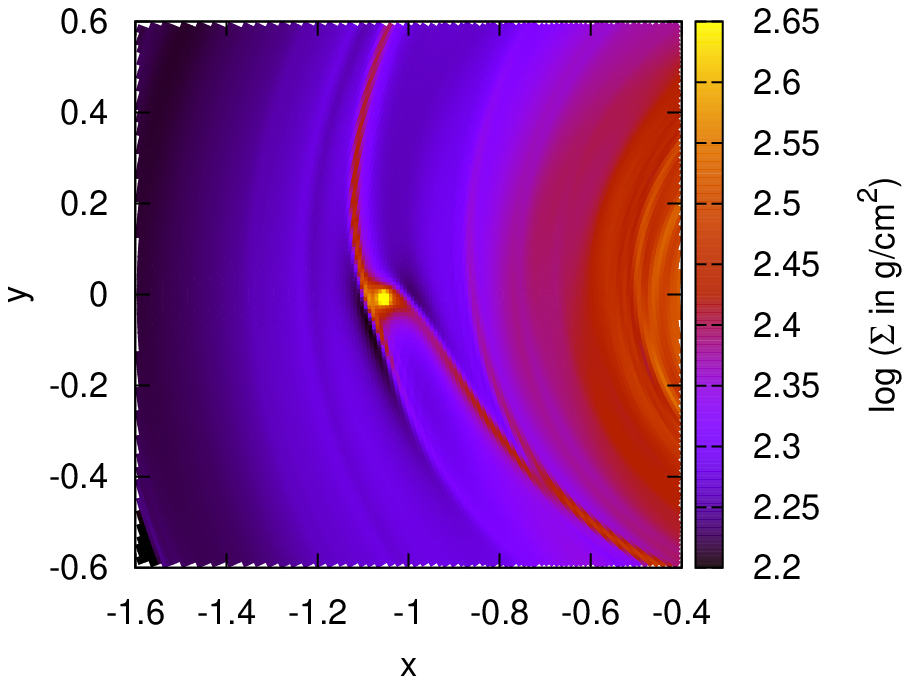}

 \includegraphics[width=0.823\linwx]{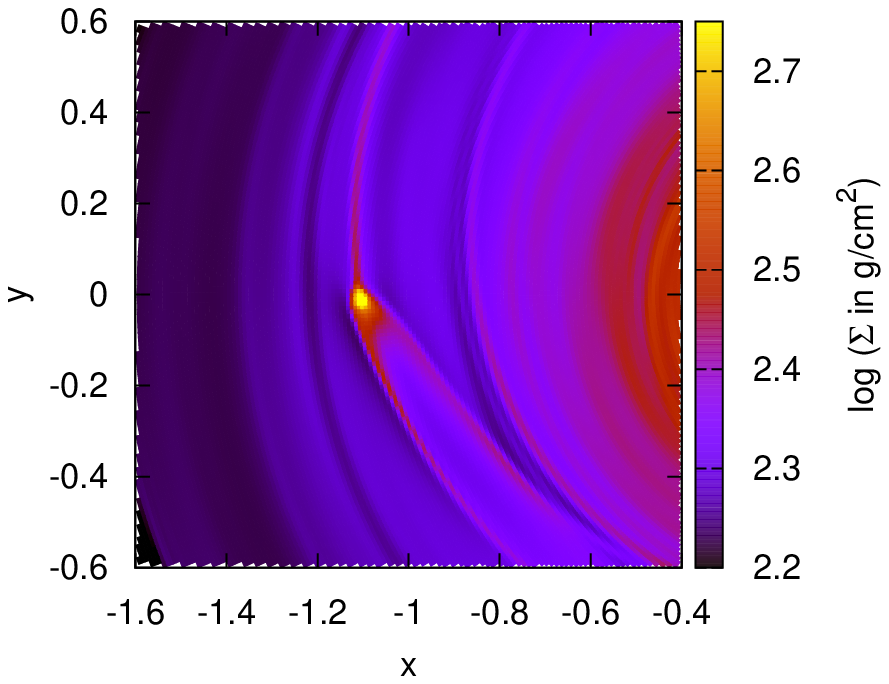}

 \includegraphics[width=0.823\linwx]{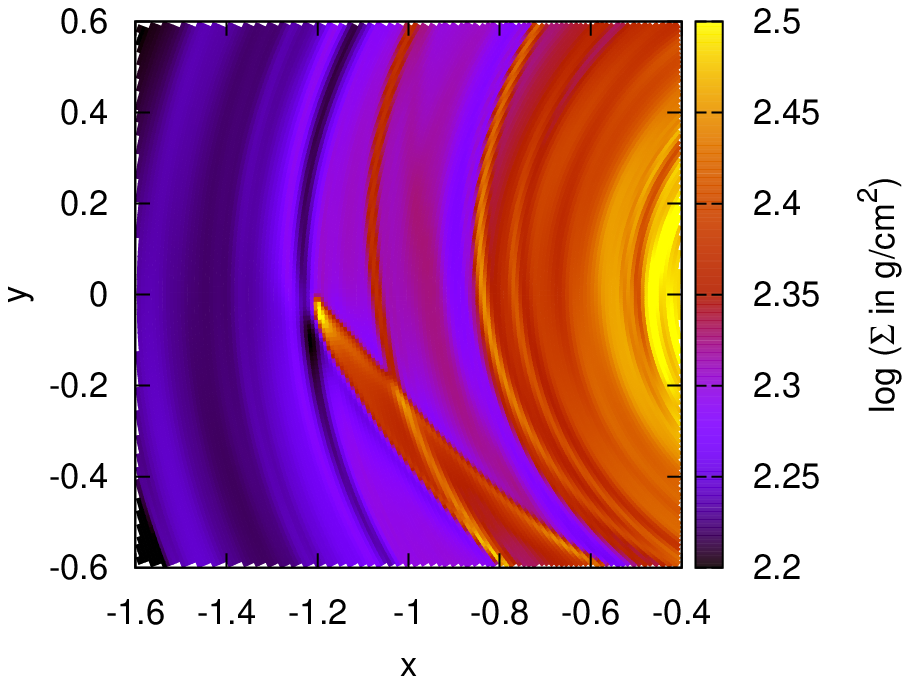}
 \caption{Displayed are the surface density maps for planets on fixed eccentric orbits at $t=150$ Orbits for isothermal simulations with $H/r=0.037$. The plots feature eccentricities of $e=0.025$, $e=0.05$, $e=0.10$ and $e=0.20$ from top to bottom. The planet is located at apoastron, meaning $(x_p,y_p) = (-(a+e),0)$, for each displayed eccentricity.
   \label{fig:IsoRhoEcc}
   }
\end{figure}

In Fig.~\ref{fig:IsoRhoEcc} we display the surface density of planets moving on fixed eccentric orbits with $e=0.025$, $e=0.05$, $e=0.10$ and $e=0.20$ in isothermal discs. These plots are taken at $t=150$ Orbits. For all surface densities displayed the planet is at apoastron, meaning the $x$ value of the planet is $-(a+e)$, with $a=1.0$, while the $y$ value of the planet is $0$ for all cases. 

Despite the fact that for eccentric orbits the density structure and flow patterns appear and disappear periodically in phase with the orbit, one can see for the $e=0.025$ case clearly two spiral waves exerted from the planet (one in the outer disc ($r > r_p$) and one in the inner disc ($r < r_p$)). The spiral wave structure is comparable to the zero eccentric case. 

\begin{figure}
 \centering
 \includegraphics[width=0.9\linwx]{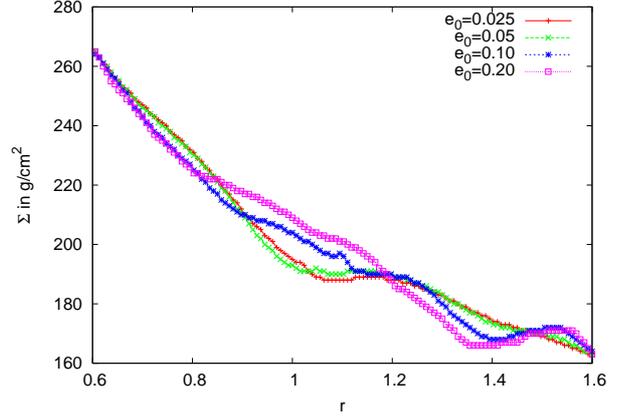}
 \caption{Azimuthally averaged surface density for planets on
  fixed eccentric orbits in the isothermal case. The displayed densities correspond to the surface
  maps displayed in the previous Fig.~\ref{fig:IsoRhoEcc}.
   \label{fig:SigIso}
   }
\end{figure}

For higher eccentricities ($e=0.05$ and $e=0.10$) the outer spiral wave is more pronounced. At the time the snapshot was taken, the planet lies in apoastron where it moves more slowly through the gas, meaning that it is overtaken by disc matter on orbits that lie both inside and outside the planetary location. We also note a significant density enhancement close to the vicinity of the planet, which lies in front of the planet and will drag the planet forward, exerting a positive torque, see Fig.~\ref{fig:EccGammaIsofull}. The reason for this phenomenon lies in the flow lines, which are distorted by the planet's gravitational potential, and come to a focus in front of the planet.
The corresponding azimuthally averaged density is displayed
in Fig.~\ref{fig:SigIso}. Interestingly, for higher eccentricities there is no gap visible anymore in the averaged
$\Sigma$-profile due to the complex structure of the spiral arms.
These are actually the reason for the high torque density displayed in Fig.~\ref{fig:EccGammatimeIsofull} (for $t=150$ orbits). At periastron this effect is reversed, leading to a negative torque acting on the planet. For $e=0.20$ we even see a stronger distortion in the density structure at apoastron. A more eccentric orbit of the planet will reduce the speed of it at apoastron more and leads thus to a more distorted density structure, compared to the zero eccentricity case.

\begin{figure}
 \centering
 \includegraphics[width=0.9\linwx]{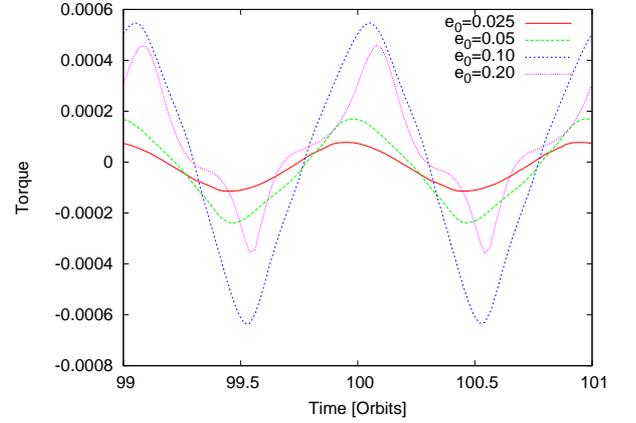}
 \caption{Torque acting on the planet during the time of $2$ planetary orbits for planets on
  fixed eccentric orbits in the isothermal case. The displayed torques correspond to the surface
  density maps displayed in Fig.~\ref{fig:IsoRhoEcc}.
   \label{fig:TztimeIso}
   }
\end{figure}

Comparing the torque density of the fully radiative simulation (see Fig.~\ref{fig:EccGammaIsofull}) with the isothermal torque density one notices only small differences for the different eccentricities. The torque acting on the planet is slightly larger for the fully radiative simulation compared for the isothermal simulation. For smaller eccentricities the reason for this phenomena lies in the spiral waves exerted from the planet. For very low eccentric planets, the arguments for a positive torque acting on the planet are the same as for planets moving on fixed circular orbits, which is explained in much detail in \citet{2009A&A...506..971K}.

\begin{figure}
 \centering
 \includegraphics[width=0.823\linwx]{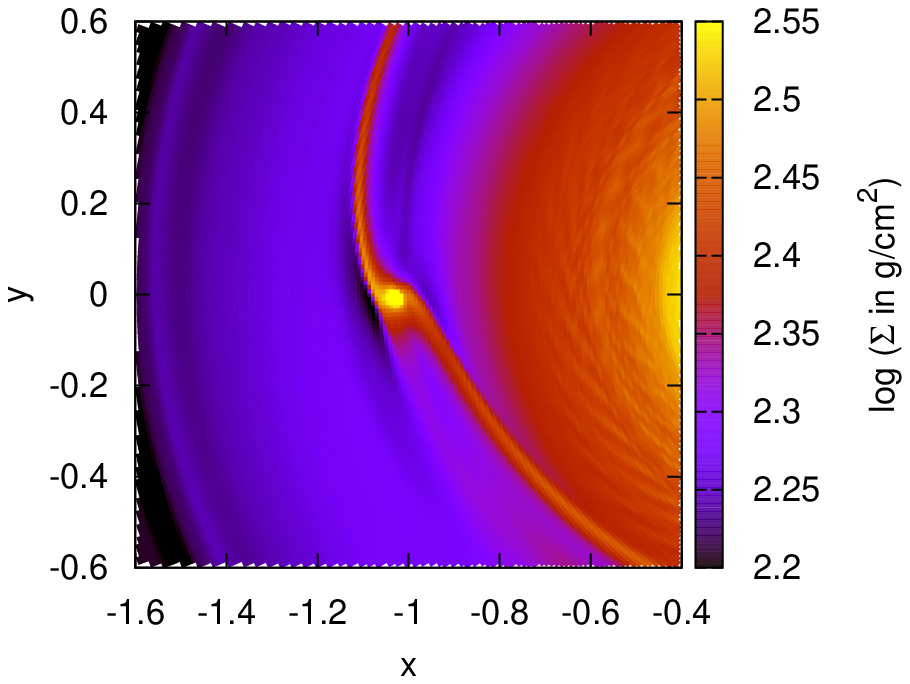}
 \includegraphics[width=0.823\linwx]{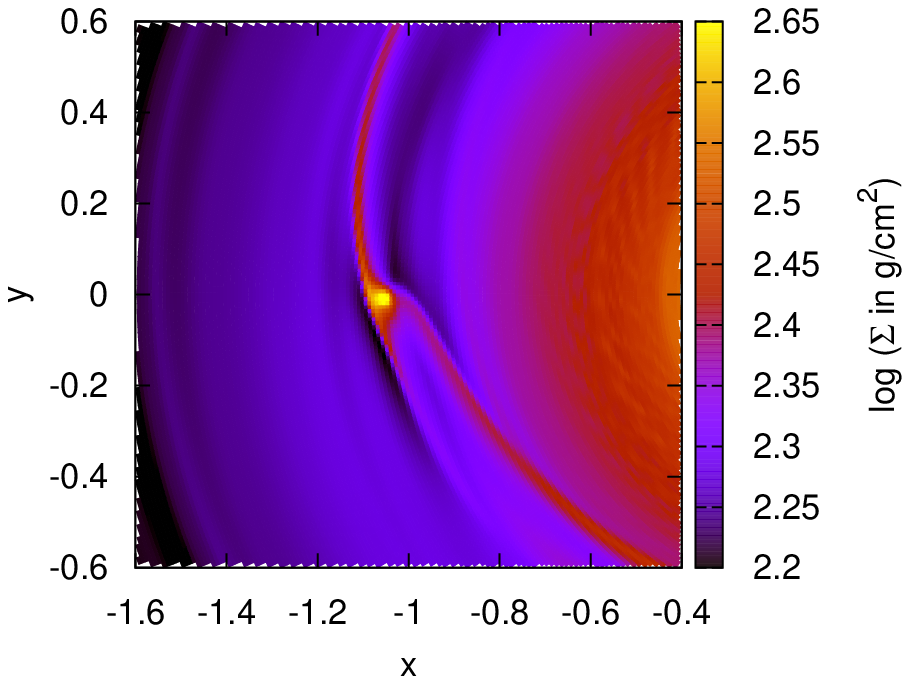}
 \includegraphics[width=0.823\linwx]{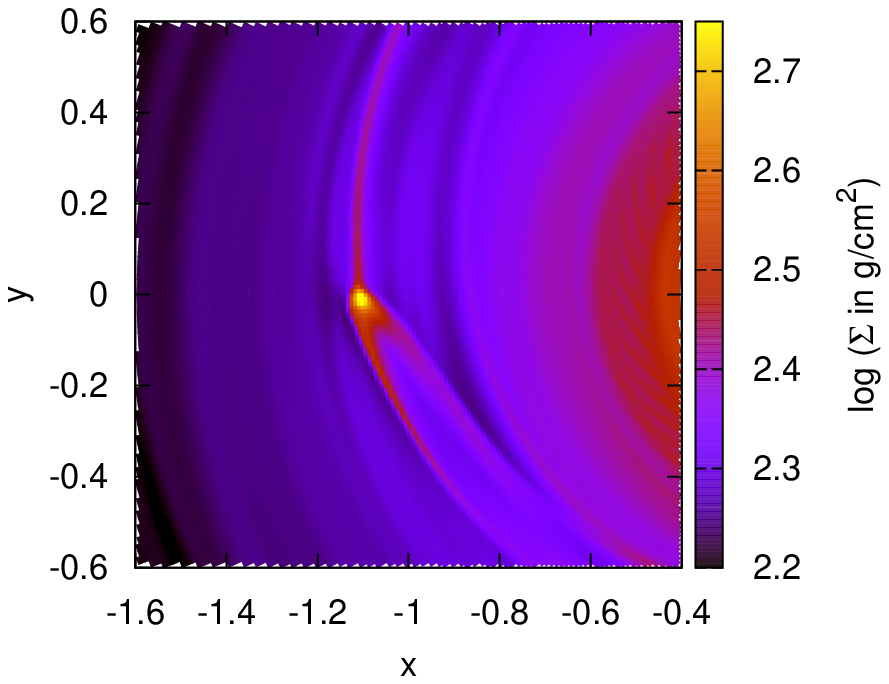}
 \includegraphics[width=0.823\linwx]{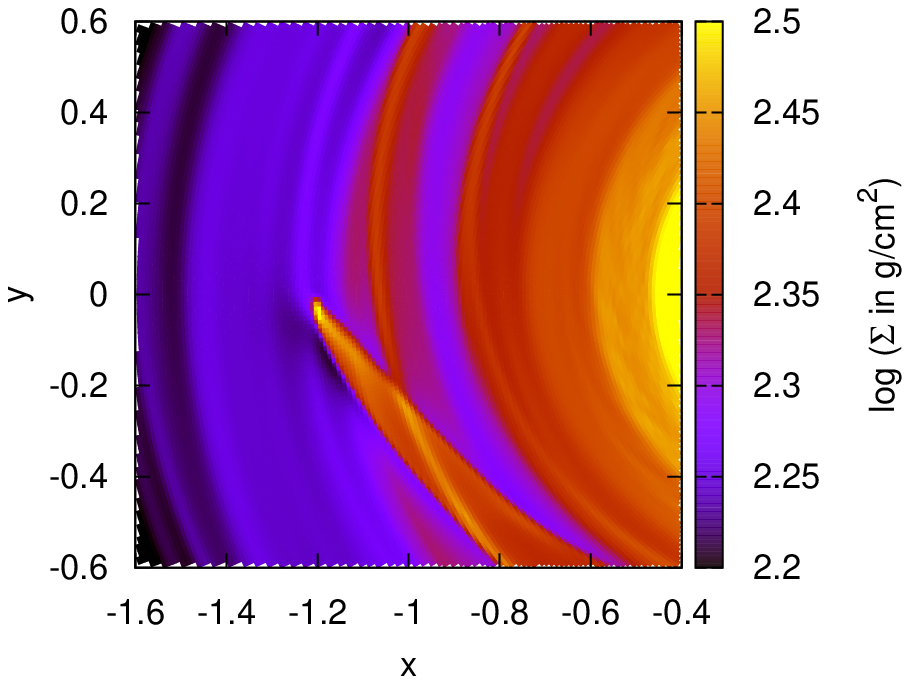}
 \caption{Displayed are the surface density maps for planets on eccentric orbits on fixed orbits at $t=100$ Orbits for fully radiative simulations. The plots feature eccentricities of $e=0.025$, $e=0.05$, $e=0.10$ and $e=0.20$ from top to bottom. The planet is located at apoastron, meaning $(x_p,y_p) = (-(a+e),0)$, for each displayed eccentricity.
   \label{fig:FullRhoEcc}
   }
\end{figure}

In Fig.~\ref{fig:EccGammatimeIsofull} we plot the change of torque density during one orbit for the fully radiative case for a planet on a fixed eccentric orbit with $e=0.1$. The orbital evolution of the torque density $\Gamma (r)$ for the fully radiative simulation does not differ much from the isothermal one. One exception is the absolute value of the torque density at the location of the planet during the time of evolution in one orbit, which is higher for the isothermal case. The change of the torque density during one planetary orbit has the same reasons as for the isothermal case.
In Fig.~\ref{fig:TztimeIso} we display the torque acting on planets on stationary eccentric orbits in the isothermal case. The eccentricities of the planet correspond to those shown in Fig.~\ref{fig:IsoRhoEcc} for the surface density. One can clearly see that planets with a higher eccentricity (to about $e \approx 0.10$) show a higher amplitude in the torque acting on the planet, but for even higher eccentricities the amplitude is being reduced. One should also keep in mind that the averaged torque acting on the planet is negative for small eccentricities while it is positive for larger ($e \geq 0.10$) eccentricities. During periapses ($t=99.5$ and $100.5$) the planet also experiences a large energy loss (not displayed here), which follows in trend the same structure as the torque profile (e.g. the energy loss is greatest for $e \approx 0.10$). This leads to an enhanced inward migration for planets with an eccentricity around $e\approx 0.10$, as can be seen in Fig.~\ref{fig:AIsofullrate} for the calculated migration rate.

%\subsection{Fully radiative discs} 

\begin{figure}
 \centering
 \includegraphics[width=0.9\linwx]{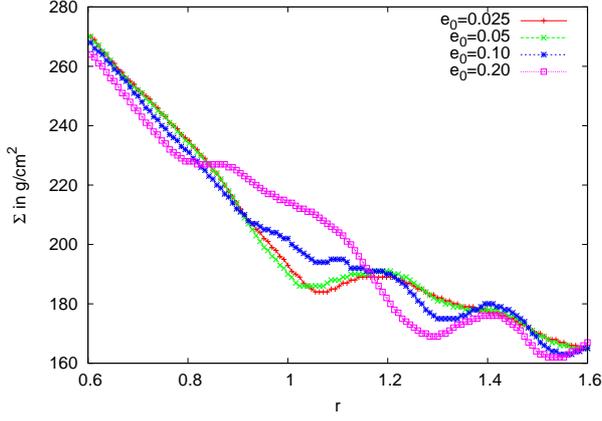}
 \caption{Azimuthally averaged surface density for planets on
  fixed eccentric orbits in the fully radiative case. The displayed densities correspond to the surface
  maps displayed in the previous Fig.~\ref{fig:FullRhoEcc}
   \label{fig:Sigfull}
   }
\end{figure}

The top figure in Fig.~\ref{fig:FullRhoEcc} displays the surface density for an $e=0.025$ planet on a fixed eccentric orbit in a fully radiative disc. The surface density distribution shown in this figure is very similar to those found for fixed circular orbits \citep[see][]{2009A&A...506..971K}. For higher eccentricities the spiral wave structure can no longer be observed as clearly as for the low eccentricity case, no matter whether we are in the isothermal or fully radiative regime. The overall surface density structure for high eccentric planets for the fully radiative case matches nearly the corresponding structure in the isothermal case. As we calculate the torque acting on the planet due to the interaction with the disc material it is not surprising that the torques acting on the planet for high eccentric planets are very similar for the isothermal and fully radiative simulations. The corresponding azimuthally averaged density is displayed
in Fig.~\ref{fig:Sigfull}. The profiles look very similar to those of the isothermal case,
where for larger eccentricities the gap becomes invisible in the 
$\Sigma$-profile.

\begin{figure}
 \centering
 \includegraphics[width=0.9\linwx]{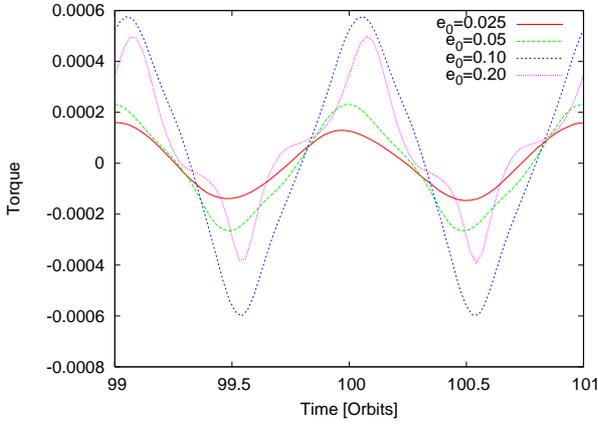}
 \caption{Torque acting on the planet during the time of $2$ planetary orbits for planets on fixed eccentric orbits in the fully radiative case. The displayed torques correspond to the surface density maps displayed in Fig.~\ref{fig:FullRhoEcc}.
   \label{fig:Tztimefull}
   }
\end{figure}

In Fig.~\ref{fig:Tztimefull} we display the torque acting on planets on stationary eccentric orbits in the fully radiative case. The eccentricities of the planet correspond to those shown in Fig.~\ref{fig:FullRhoEcc} for the surface density. One can clearly see that planets with a higher initial eccentricity (to about $e \approx 0.10$) show a higher amplitude in the torque acting on the planet, but for even higher eccentricities the amplitude is being reduced. One should also keep in mind that the averaged torque acting on the planet is positive for very small eccentricities ($e \leq 0.025$) and negative for $0.025 < e < 0.10$ while it is positive for larger ($e \geq 0.10$) eccentricities. During periapses ($t=99.5$ and $100.5$) the planet also experiences a large energy loss (not displayed here), which follows in trend the same structure as the torque profile (e.g. the energy loss is greatest for $e \approx 0.10$). This leads to an enhanced inward migration for planets with an eccentricity around $e\approx 0.12$, as can be seen in Fig.~\ref{fig:AIsofullrate} for the calculated migration rate.

%For high eccentric planets the spiral wave structure can no longer be observed for both cases, but as the trend for the torque acting on the planet is positive for both kind of simulations (with only a small differences in the total value) the inclusion of radiation transport/cooling for high eccentric orbits is no longer important. The inclusion of radiation transport/cooling has its major effects for planets with a low eccentricity, because the spiral wave structure for these kind of orbits is more similar to the circular one.

\section{Moving planets on initial eccentric orbits}

To study eccentricity damping of a planet embedded in a protoplanetary disc dynamically we now change our configuration in such a way, that the planet is able to evolve its orbit in time. In the beginning of the simulation we put the planet in the disc, and let the mass of the planet grow in such a way that the planet reaches its final mass of $m_{p} = 20 M_{Earth}$ at $t=10$ orbits. This way we do not disturb the density distribution of the disc as much as by putting the planet with its full mass immediately into the unperturbed disc. 

\subsection{Isothermal discs}

\begin{figure}
 \centering
 \includegraphics[width=0.9\linwx]{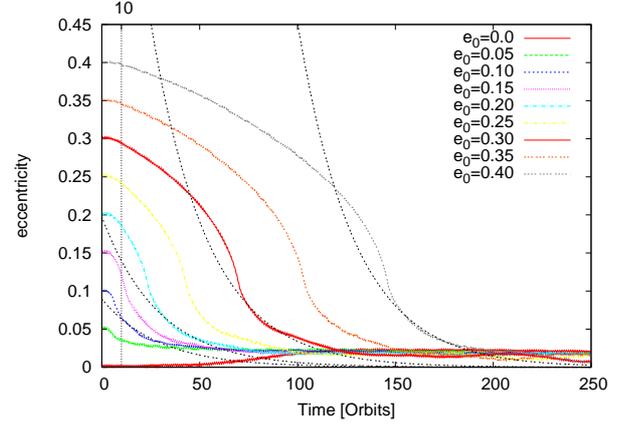}
 \caption{Time evolution of eccentricity of planets embedded in isothermal discs with $H/r=0.037$ and individual starting eccentricity. 
 The vertical dotted line at $t=10$ indicates the time at which the planets have grown to their full mass.
 The black dashed lines indicate an exponential decay with a fitted $\tau_{ecc}=29$ shifted to obtain a match for $e_0=0.1, 0.2, 0.3$ and $0.4$. 
   \label{fig:EccIsoEccall}
   }
\end{figure}

The eccentricity evolution of various planets with individual starting eccentricity can be seen in Fig.~\ref{fig:EccIsoEccall}. Planets with an initial eccentricity lower than about $e \approx 0.10$ experience an exponential damping of eccentricity (as soon as the planet has reached its destined mass), while planets with larger initial eccentricity adopt initially a slower damping. As soon as planets with a higher initial eccentricity reach an eccentricity of about $e \approx 0.10$ they undergo an exponential damping of eccentricity as well. For eccentricities $0.10 < e < 0.15$ the damping of eccentricity is accelerated compared to the damping for higher eccentricities. This was expected as a result of our calculations of the change of eccentricities for planets on fixed eccentric orbits (see Fig.~\ref{fig:EccIsofullrate}). In the end all planets have settled to about the same value of eccentricity ($e\approx 0.02$), independent of their starting eccentricity.

Using linear analysis for small eccentricities \citet{2004ApJ...602..388T} find that the mean eccentricity change (averaged over one planetary orbit) is given by

\begin{equation}
	\frac{\overline{de/dt}}{e} = - \frac{0.780}{t_{wave}}
\end{equation}
with the characteristic time
\begin{equation}
	t_{wave}=q^{-1} \left(\frac{\Sigma_P a^2}{M_\ast}\right)^{-1} \left(\frac{c_s}{a\Omega_P}\right)^4 \Omega_P^{-1},
\end{equation}
where $q$ denotes the mass ration between the planet and the star and $\Sigma_P$ the local surface density at the planetary orbit. For our $20 M_{Earth}$ planet at $5.2 \AU$ in the isothermal $H/r=0.037$ disc we find the characteristic time to be $t_{wave}=7.70$ orbits, which gives an eccentricity damping time scale of about $\tau_{ecc} = t_{wave}/0.78 = 9.88$ orbits. Apparently, this theoretically estimated decay time scale does not match the fitted value of $\tau_{ecc} = 29$
as obtained from our numerical results. This exponential decay time is indicated by the black dashed lines in Fig.~\ref{fig:EccIsoEccall}.
As will be seen below, this strong difference occurs only for the isothermal case and is much reduced in the fully radiative models.
In previous simulations on the evolution of eccentric planets {\it in isothermal discs} the agreement between theory and numerics has been much better
\citep{2007A&A...473..329C}. We attribute the difference to two effects, a smaller $H/r$ and the usage of the steeper more realistic cubic-potential
instead of the more shallow $\epsilon$-potential that was used in \citet{2007A&A...473..329C}.
Indeed, an additional run for the $e_0=0.20$ case with the $\epsilon$ potential using $r_{sm} =0.8$ yields a fitting parameter of $\tau_{ecc} = 17$.
This is much closer to the theoretical computed value. 
%%On the other hand, this result is still about a factor of $2$ larger than the theoretical result, which indicates that the resolution in the isothermal case for moving planets in discs with this small aspect ration of $H/r=0.037$ has to become much larger.

\begin{figure}
 \centering
 \includegraphics[width=0.9\linwx]{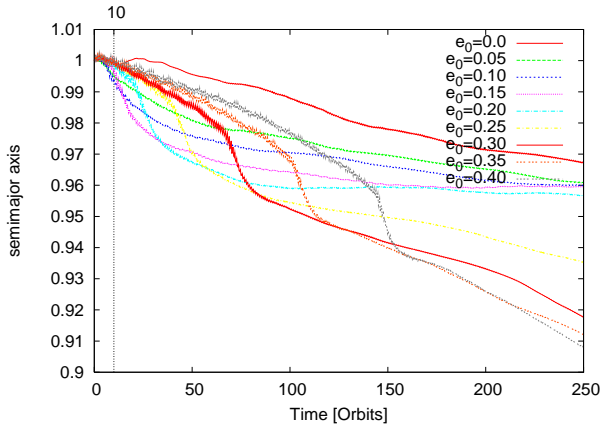}
 \caption{Time evolution of the semi-major axis of planets embedded in isothermal discs with $H/r=0.037$ and individual starting eccentricity.
   \label{fig:AIsoEccall}
   }
\end{figure}

But not only the eccentricity evolves in time on a moving planet, but also the semi-major axis of the planet as can be seen in Fig.~\ref{fig:AIsoEccall}. The reduction of the semi-major axis is in direct correlation with the damping of eccentricity. For planets with initially low eccentricity (lower than $e=0.12$), we see a immediately reduction of the semi-major axis on a fast rate. This rapid inward migration is then reduced to a very slow migration rate when the planets reach their final value of eccentricity ($e\approx 0.02$). Planets with higher initial eccentricity first migrate inward at a slower rate as their low eccentric counterparts, but at the time their eccentricity reaches the above mentioned point of $e\approx 0.10$ their migration rate changes and they undergo a rapid inward migration, as their eccentricity damps exponentially. This rise in the migration rate was actually expected from the calculations of the theoretical migration rate for planets on fixed eccentric orbits (see Fig.~\ref{fig:AIsofullrate}). When the eccentricity damping then changes to a slower decay the migration rate becomes nearly linear and the planets move inward with a constant rate again. The planet starting with a zero eccentricity attains a low eccentricity during time and is also the planet migrating inward with the slowest speed. These results lead to the conclusion that planets on eccentric orbits do migrate inward at a slightly faster speed compared to their circular counterparts and that planets starting from initial higher eccentricity migrate inward faster than those with a lower initial eccentricity.

\begin{figure}
 \centering
 \includegraphics[width=0.9\linwx]{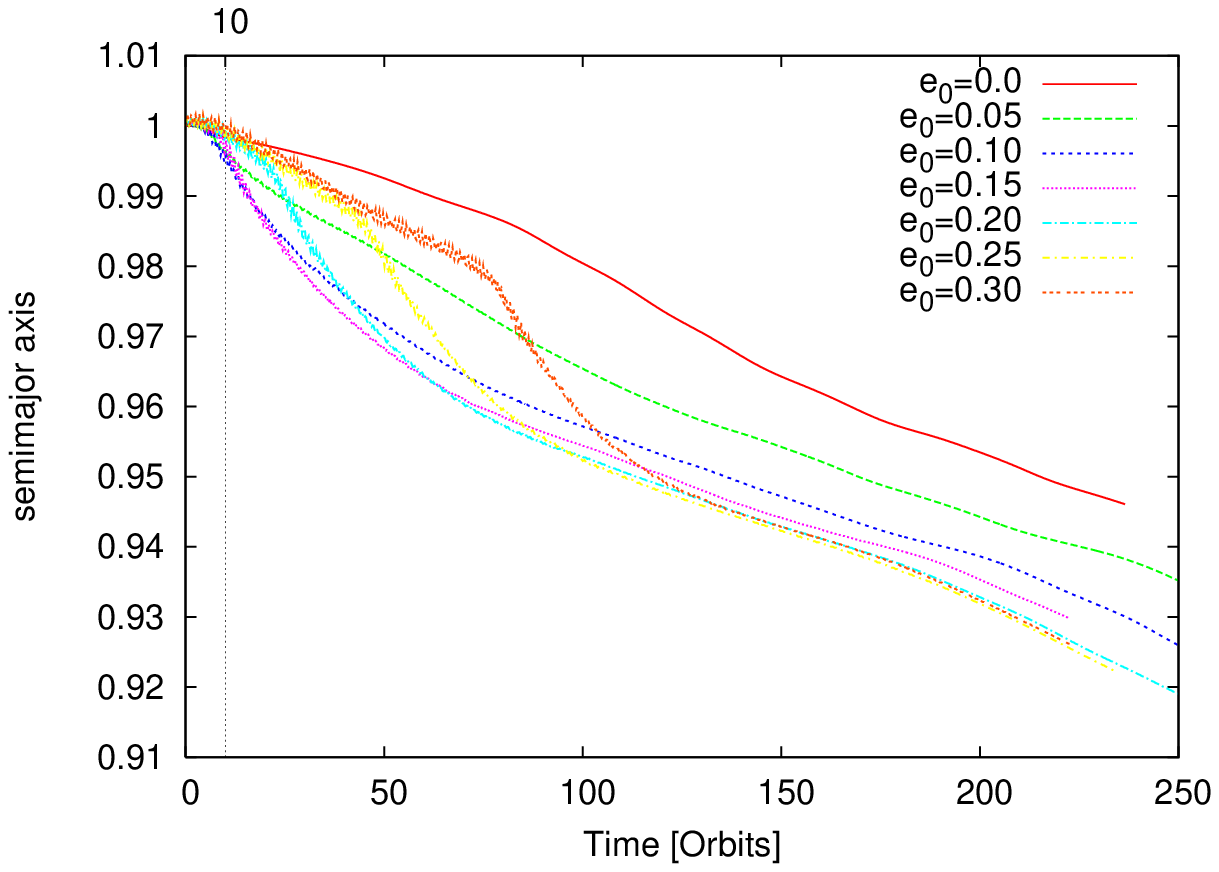}
 \includegraphics[width=0.9\linwx]{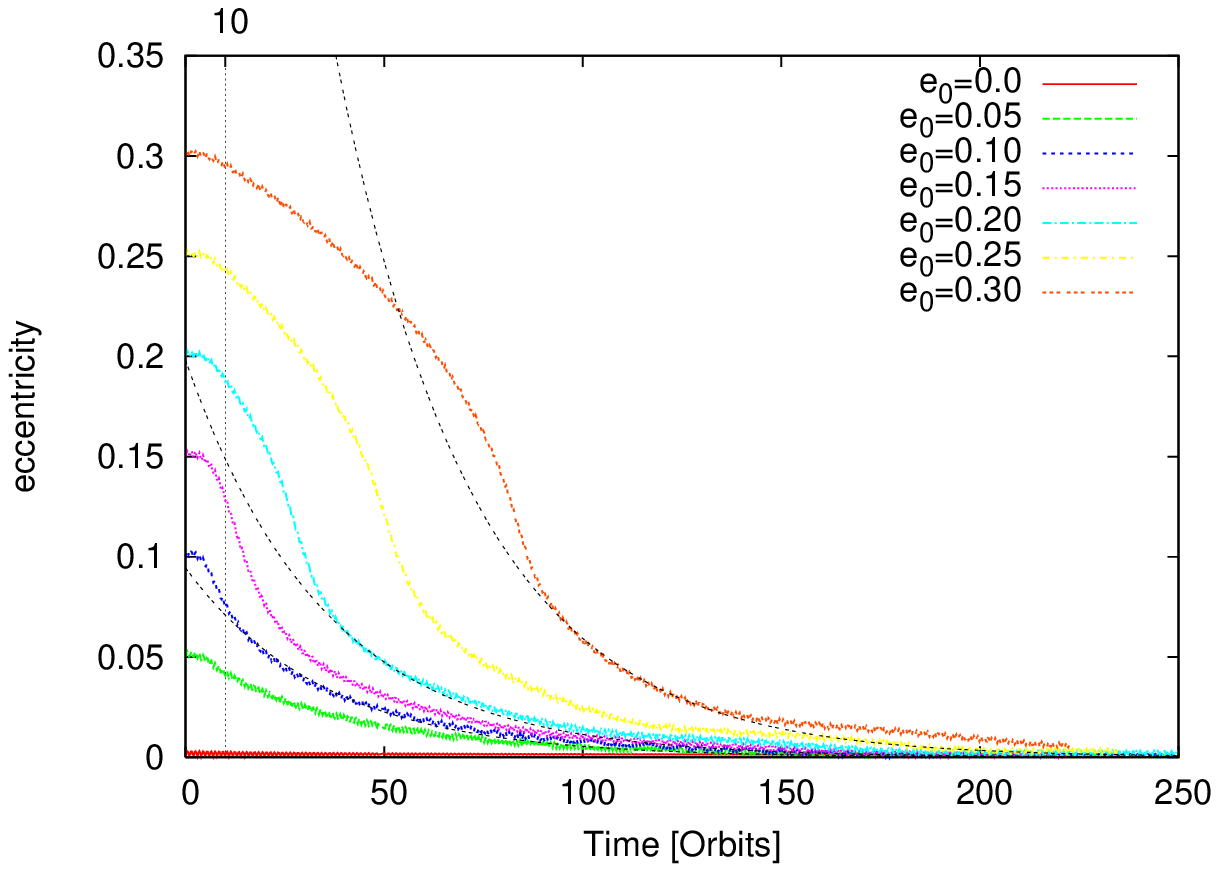}
 \caption{Time evolution of the semi-major axis (top) and eccentricity (bottom) of a $20 M_{Earth}$ planet in an isothermal $H/r=0.05$ disc. The black dashed lines  indicate an exponential decay of eccentricity with $\tau_{ecc}=35$ for $e_0=0.1, 0.2$ and $0.3$.
%%  This exponential decay does not match our calculated $\tau_{ecc} = 32.95$ exactly, but is within good agreement.
   \label{fig:AExcIsoHr005}
   }
\end{figure}

But not only the initial eccentricity of the planet changes the evolution of the planet, but also the disc parameters. Our above shown simulations used a $H/r=0.037$ in the isothermal case. We now compare these results with isothermal simulations featuring $H/r=0.05$. In Fig.~\ref{fig:AExcIsoHr005} we display the change of the semi-major axis and the eccentricity of $20 M_{Earth}$ planets in an isothermal $H/r=0.05$ disc. The other disc parameters are the same as for the $H/r=0.037$ planets. On the one hand the overall trend that the planet starting from a zero eccentricity orbit has the slowest inward migration is also true for the simulations with $H/r=0.05$. On the other hand the inward migration in a $H/r=0.05$ disc is faster for all initial eccentricities. In a $H/r=0.037$ disc the planets with an initial eccentricity lower than $e_0 =0.20$ end up with a very slow inward migration, while in the $H/r=0.05$ case all planets end up with nearly the same migration rate, even the planets with an initial high eccentricity. These findings are in very good agreement with the results of \citet{2007A&A...473..329C}.

From our disc-data we estimate a theoretical exponential decay time for eccentricity damping timescale as $\tau_{ecc}=32.95$ orbits. 
This value is larger than the previous one for the lower $H/r=0.037$ case by a factor of $(0.05/0.037)^4$.
Numerically we obtain  $\tau_{ecc} = 35$, see the black dashed curves in Fig.~\ref{fig:AExcIsoHr005}. 
Hence, the numerical exponential decay in the simulations matches quite well to the theoretical damping rate. We attribute this better match to the 
smoothing effect of the higher pressure in the disc.

%%
%%When we now compare the eccentricity of the planets in a $H/r=0.05$ disc to a $H/r=0.037$ disc we note two major differences. In the $H/r=0.05$ case the eccentricity damping is slower and it settles down to nearly zero eccentricity. A higher aspect ratio of the disc leads to a density structure in which more material is located above and below the planet compared to a disc with a lower aspect ration. The material above and below the planet prevents the planet to form a gap (or even a partial gap) in the disc, which will change its migration rate.
%%

\subsection{Fully radiative discs}

\begin{figure}
 \centering
 \includegraphics[width=0.9\linwx]{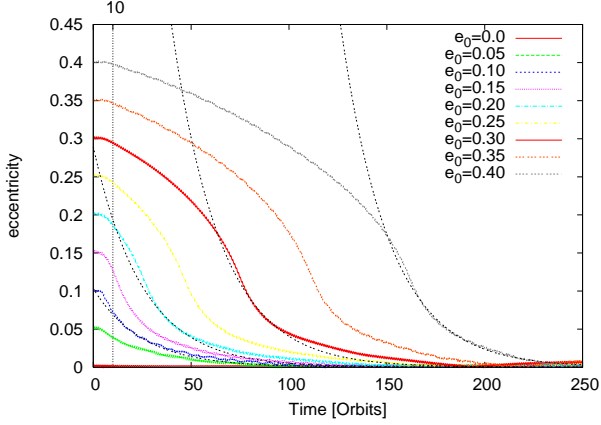}
 \caption{Time evolution of eccentricity for planets with individual starting eccentricity for the fully radiative case. The black dashed lines indicate an exponential decay with $\tau_{ecc}=25$  for $e_0=0.1, 0.2, 0.3$ and $0.4$. 
  %% {\bf This value is about $25\%$ of from the calculated value of $\tau_{ecc}=19.37$.}
   \label{fig:EccfullEccall}
   }
\end{figure}

\begin{figure}
 \centering
 \includegraphics[width=0.9\linwx]{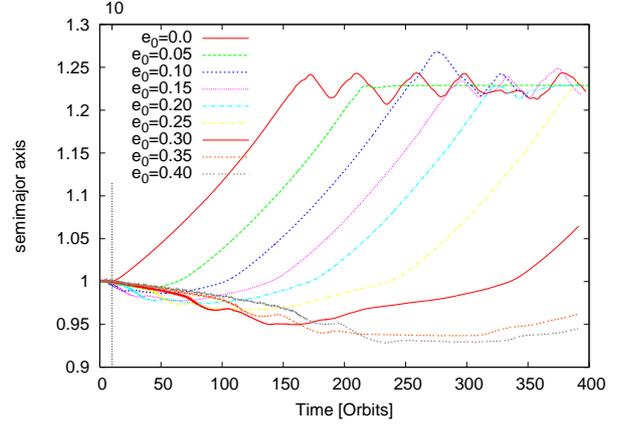}
 \caption{Time evolution of the semi-major axis of planets with individual starting eccentricity for the fully radiative case for $20 M_{Earth}$ planets.
   \label{fig:AfullEccall}
   }
\end{figure}

If we now let the planet evolve its orbit with time for the fully radiative case, we obtain a very similar behaviour to the isothermal case for the evolution of eccentricity (see Fig.~\ref{fig:EccfullEccall}), but a quite different behaviour concerning the semi-major axis evolution (see Fig.~\ref{fig:AfullEccall}).  The damping of eccentricity for the fully radiative discs proceeds on a comparable timescale to the
two previous results, and we find for the exponential behaviour $\tau_{ecc}=25$. 

The inclusion of radiation transport/cooling seems to have only a little effect on the damping of eccentricity. The damping of eccentricity is somewhat slower in the fully radiative case compared to the isothermal simulations. For planets with eccentricities lower than about $e=0.10$ the damping of eccentricity follows approximately an exponential law. The eccentricity damping for high eccentric planets is first slower, until the eccentricity reaches a value of about $e=0.10$ and is then increased to an exponential value (see the fit in Fig.~\ref{fig:EccfullEccall}). This rise in the speed of eccentricity damping is expected from the calculation of the theoretical change of the migration rate for planets on fixed eccentric orbits (see Fig.~\ref{fig:EccIsofullrate}).  The reduction rate of eccentricity slows down when it reaches $e\approx 0.05$. In the end, as for the isothermal case, the eccentricity reaches the same value for all starting eccentricities, but the absolute value of eccentricity for the isothermal case ($e\approx 0.02$) is about a factor of five higher as for the fully radiative case ($e\approx 0.004$). The characteristic time for eccentricity damping in our fully radiative disc for the embedded $20 M_{Earth}$ planet is $t_{wave}=15.1$ orbits, leading to a eccentricity damping time scale of $\tau_{ecc}=19.37$ orbits, if we use $H/r=0.037$ and an adiabatic sound speed. The black dashed lines in Fig.~\ref{fig:EccfullEccall} indicate the exponential decay for our fitting, $\tau_{ecc} = 25$.
This is not exactly matching our theoretical results but the agreement is satisfactory.
Note that as soon as the eccentricity of our simulations reaches $e\approx 0.05$ we observe a difference between our numerical results and the theoretical
eccentricity damping. The damping in the simulations is much slower than the theoretical damping, as the planet stops inward migration at this point and stays on an orbit with nearly constant semi-major axis, which might slow down eccentricity damping.

The density structure near the planet is smoothed for the fully radiative case. As the mass in the disc and near the planet is able to cool, it will give a smoother density profile. If the eccentricity is damped to a small enough value, the effects of radiation transport/cooling can set in, and will give rise to a positive torque, which results in outward migration (see plot of the semi-major axis Fig.~\ref{fig:AfullEccall}). 
As long as the eccentricity of the planet is higher than $\approx 0.03$ the planet will migrate inward, even for the fully radiative simulations, but with a rate smaller than the corresponding isothermal simulations. When the eccentricity reaches the critical value of $e \approx 0.10$ we observe a bump in the semi major axis of the planet. We expected this as the migration rate determined by planets on fixed eccentric orbits (see Fig.~\ref{fig:AIsofullrate}) has its minimum there.

Starting from a zero eccentricity planet, which will not gain much eccentricity during its evolution, one can see direct outward migration, as predicted in our paper \citep{2009A&A...506..971K} for planets moving on circular orbits. Planets starting from a higher initial eccentricity will have to damp their eccentricity to a very low value to feel a positive torque acting on them and thus leading them to outward migration. The eccentricity needed for outward migration is the same for all planets in our simulations (independent of the starting eccentricity) with a average value 
$e\approx 0.03$ for reversal. As soon as the $e$-damping is complete the planet experiences a positive torque and will migrate outward, even for the highest initial eccentricity in our simulations.

%However, as the planet migrates outward in the disc it reaches regions where the discs density and temperature is reduced compared to the initial location of the planet. This reduction in density and temperature has a direct effect on the migration rate of the planet. When the planet reaches a region where the density and temperature have changed considerably for the conditions for outward migration to hold the planet will come to a halt.
%This can be clearly see in Fig.~\ref{fig:AfullEccall} for the initially low eccentric planets. This indicates the outward migration due to radiative transport/cooling will not go on forever but depends on the local disc structure.

%On the contrary we monitored the evolution of a $20 M_{Earth}$ planet starting with an initial eccentricity of zero, but at $r=1.5$ for the same disc parameters in the fully radiative scheme. The results of this simulation have shown that the planet will first start to migrate inward until it reaches the same barrier at $r=1.23$ that the planets starting at $r=1.0$ encountered. However, this time the planet stops its inward migration and stays on its orbit at about $r=1.23$.
%This result supports the suggested limit for outward migration due to the disc properties.

The inclusion of radiation transfer/cooling in disc for embedded low mass planets on eccentric orbits will result in outward migration as soon as the eccentricity of the planet is damped to a value near the circular eccentricity. If the planet reaches low eccentric orbits which are nearly circular, the density structure will also become nearly like the one of a planet on a circular orbit. The planet will then migrate outward as if it was on a circular orbit. This outward migration will then be stopped as soon as the planet reaches regions in the disc where the density and temperature are too low to support outward migration.

Looking at Fig.~\ref{fig:AfullEccall} it seems, that the outward migration is stopped for all planets at a certain critical radius in the disc. 
Even though at some point one might expect a termination of the outward migration depending on the local disc conditions,
we note that in our case this feature appears to
be a result of insufficient numerical resolution. For more information about this, see App.~\ref{app:Numfeature}.

\section{Higher mass planets on eccentric orbits}

As was shown in many previous works the migration rate of a planet embedded in a protoplanetary disc does not only depend on the disc's structure and thermodynamics, but also on the planetary mass \citep[e.g.][]{2009A&A...506..971K}. In our previous work we found that planets up to about $33 M_{Earth}$ experience a positive torque (which indicates outward migration), while higher mass planets experience a negative torque (indicating inward migration). It is now very interesting to investigate the evolution of planets with higher masses on eccentric orbits. In this chapter we focus on the eccentricity change and migration of planets ranging for $30$ to $200$ earth masses in the isothermal and fully radiative regime. We focus here directly on migrating planets, and we do not provide a torque analysis as we did for the $20 M_{Earth}$ planet.

\subsection{Isothermal discs}

In the isothermal regime using the cubic $r_{sm}=0.5$ potential may cause numerical problems (as described above) which become even
more severe when the moving planets have a higher mass. The potential becomes just too deep for higher mass planets to consider our results as correct for the grid resolution we use for moving planets. A deeper planetary potential will give rise to a much higher density distribution near the planet which will become unrealistic in the isothermal case as the disc is not able to heat up upon compression. So we use in this section the common $\epsilon$-Potential with $r_{sm}=0.8$ for planets with higher mass (as described in \citet{2009A&A...506..971K}), which will give us smoother and more realistic results in this case.
Again, we let the planet reach its final mass during a time of $10$ orbits. By this way the disturbances in the disc are not as big as by inserting the planet with its full mass at once. As the mass of the inserted planets becomes higher this feature becomes more and more important.

\begin{figure}
 \centering
 \includegraphics[width=0.90\linwx]{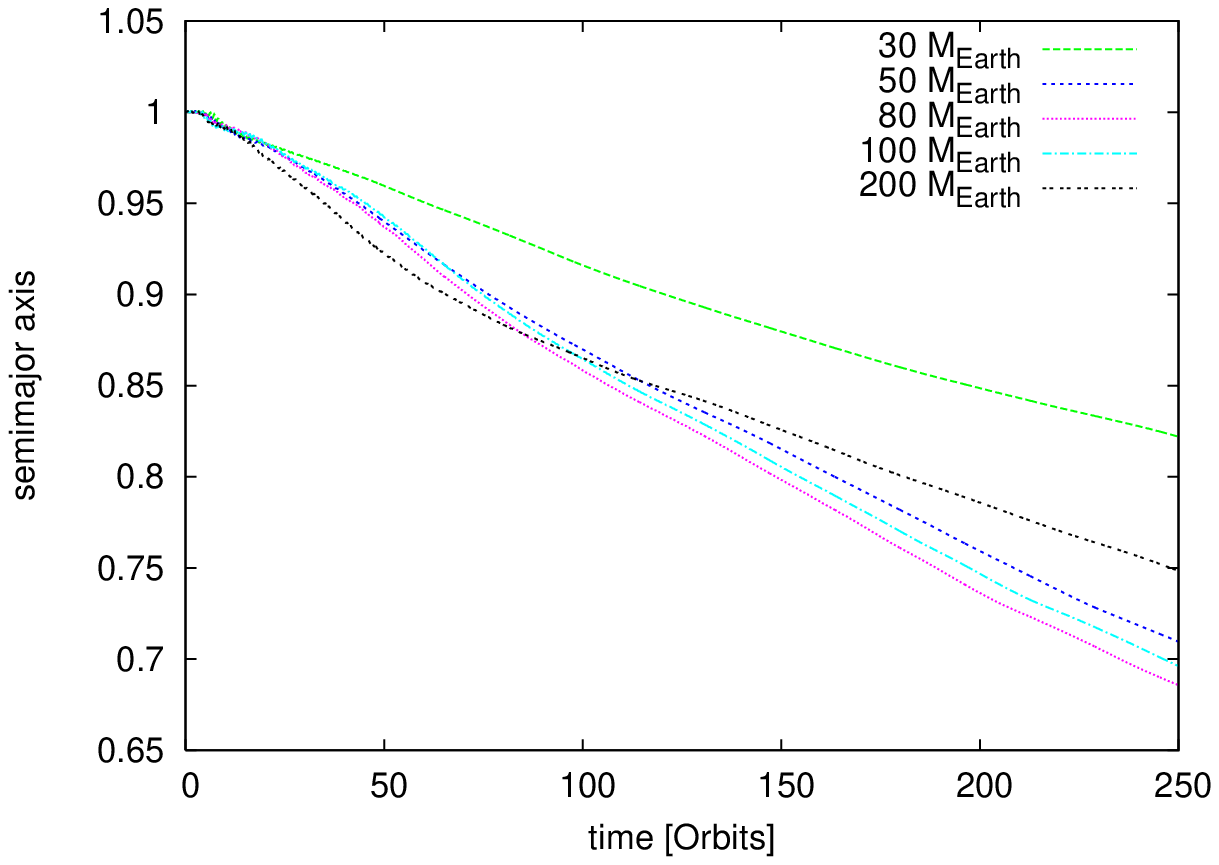}
 \includegraphics[width=0.90\linwx]{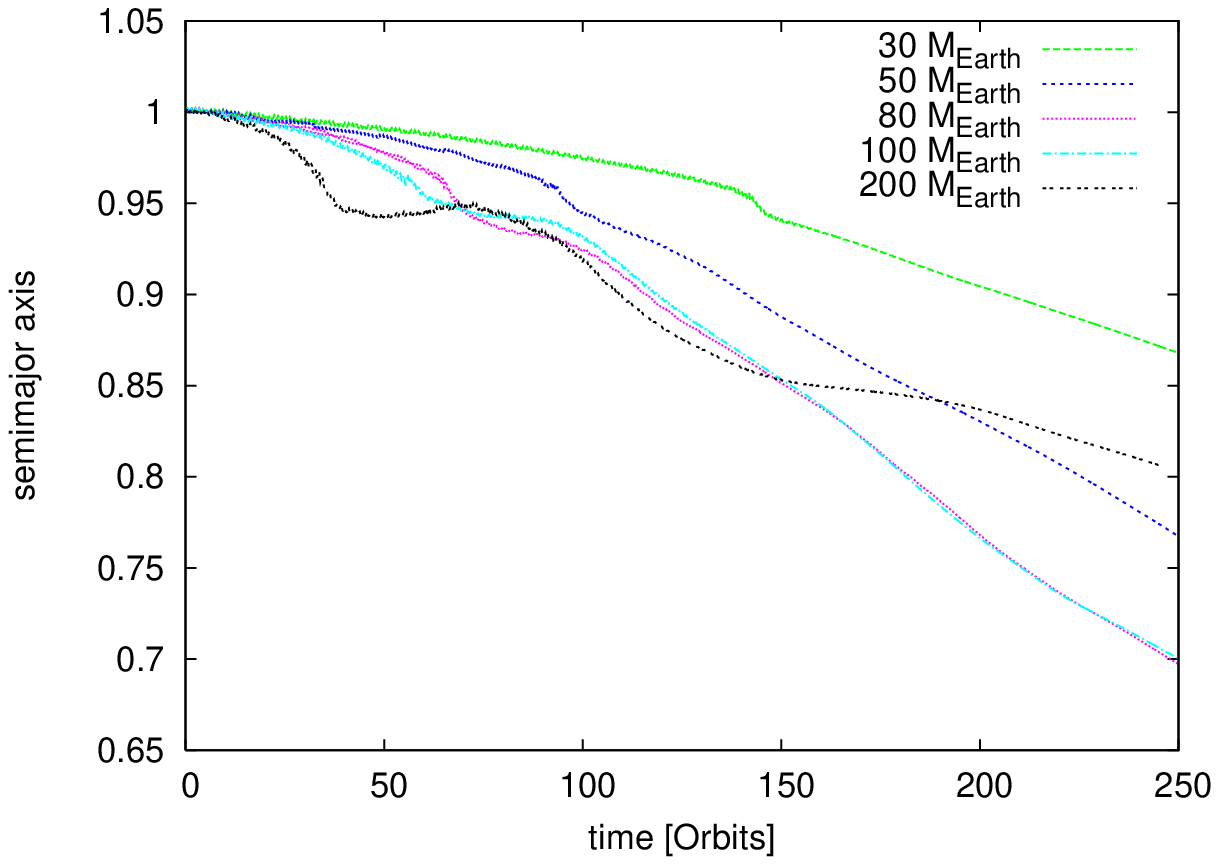}
 \caption{Time evolution of the semi-major axis for planets in an isothermal disc ($H/r=0.037$) with $30$, $50$, $80$, $100$ and $200$ Earth masses. In the top graph the planets have an initial eccentricity of $e_0=0.10$ and in the bottom plot the initial eccentricity is $e_0=0.40$.
   \label{fig:AIsoMassAll}
   }
\end{figure}

\begin{figure}
 \centering
 \includegraphics[width=0.90\linwx]{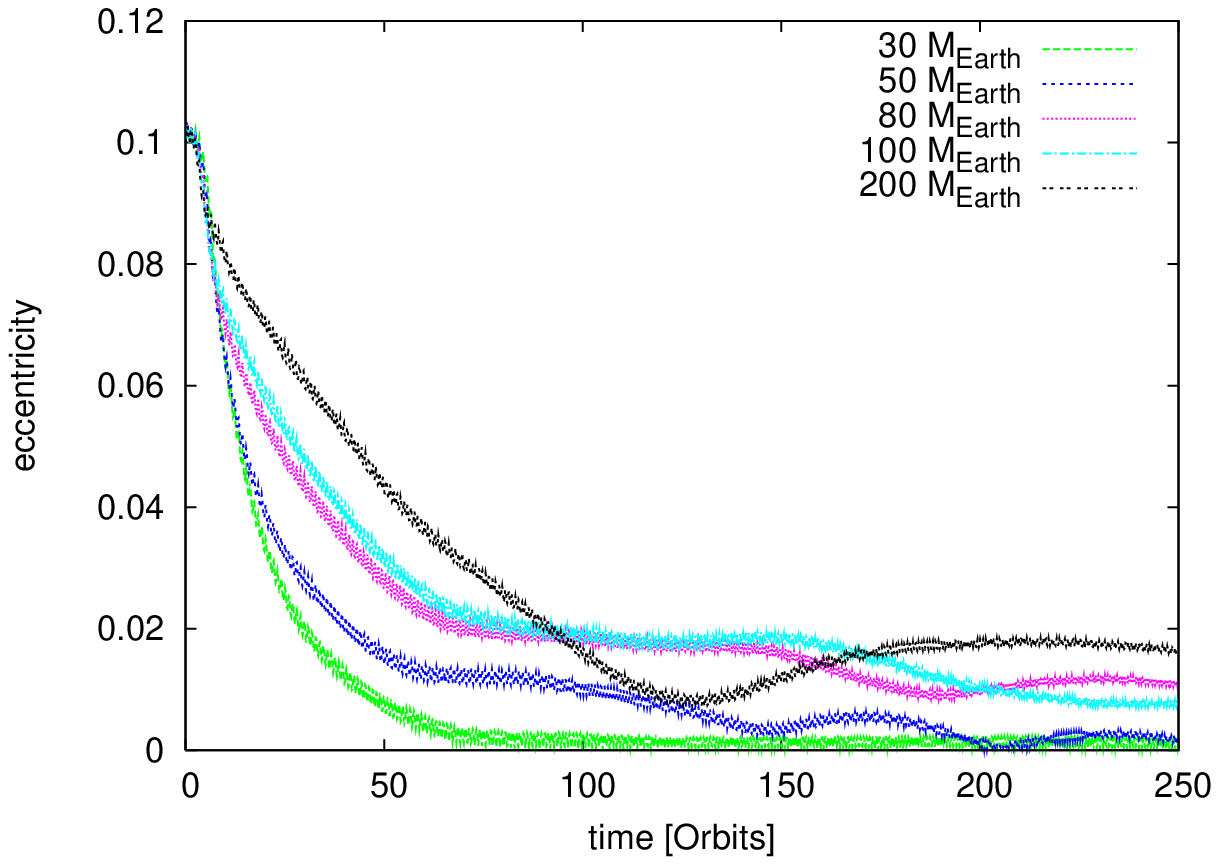}
 \includegraphics[width=0.90\linwx]{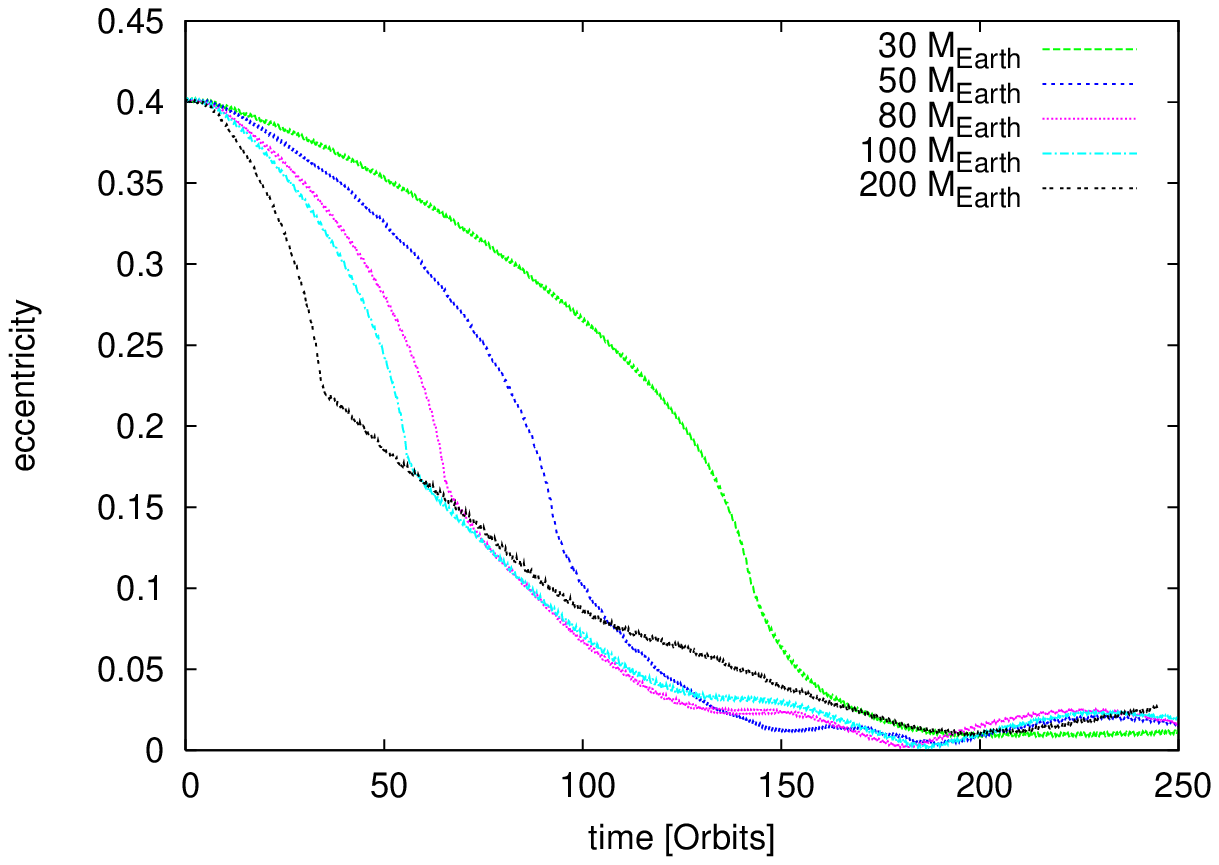}
 \caption{Time evolution of the eccentricity for planets in an isothermal disc ($H/r=0.037$) with $30$, $50$, $80$, $100$ and $200$ Earth masses. In the top figure the planets have an initial eccentricity of $e_0=0.10$ and in the bottom figure the initial eccentricity is $e_0=0.40$.
   \label{fig:ExcIsoMassAll}
   }
\end{figure}

In Fig.~\ref{fig:AIsoMassAll} we display the evolution of the semi-major axis for planets with $30$, $50$, $80$, $100$ and $200 M_{Earth}$ in the isothermal case for $e_0=0.10$ and $e_0=0.40$. The high mass planets ($M \geq 50 M_{Earth}$) migrate inward at the same rate, in contrast to the $30 M_{Earth}$ planet, when starting with $e_0=0.10$. The planets seem all to migrate inward on a linear scale. However when the planets start with $e_0=0.40$ we observe a different picture; now only the $80$, $100$ and $200 M_{Earth}$ planets migrate inward at the same speed and faster than the planets with lower mass. The migration speed is faster in the beginning for the planets starting with $e_0=0.10$ compared to those starting with $e_0=0.40$, meaning that a high initial eccentricity tends to slow down the initial inward migration for planets with a higher mass ($M \geq 30 M_{Earth}$) which is in contrast to the results found for the $20 M_{Earth}$ planet. As the eccentricity is damped during time the migration speed for the planets starting with $e_0=0.4$ becomes faster than for the $e_0=0.1$ planets. The observed bumps in the evolution of the semi-major axis occur for all planetary masses at the same time when we observe a change in the damping of eccentricity (displayed in the bottom figure of Fig.~\ref{fig:ExcIsoMassAll}).

In Fig.~\ref{fig:ExcIsoMassAll} we display the time evolution of eccentricity for planets starting with $e_0=0.10$ and $e_0=0.40$. The eccentricity drops immediately at the start of the simulation for the $e_0=0.10$ case for all planetary masses, but the damping seems slowest for planets with higher mass, and in the end all high mass planets ($M > 50$) end up with an eccentricity below $e=0.02$. In the $e_0=0.4$ case the eccentricity damping sets in as soon as the planets have reached their final mass (after $10$ orbits). The $200 M_{Earth}$ planet experiences initially the fastest eccentricity damping until $e \approx 0.22$ and then the damping is slowed down to a nearly linear damping. The smaller planets follow in principal the same trend in eccentricity damping, only that the initial damping is slower compared to the $200 M_{Earth}$ planet and terminates at a lower eccentricity. After the eccentricity reaches $e\leq 0.17$ the $100$ and $200 M_{Earth}$ planet have a very similar subsequent eccentricity and semi-major evolution; at $e\leq 0.15$ the $80 M_{Earth}$ planet joins this evolution. 

\begin{figure}
 \centering
 \includegraphics[width=0.90\linwx]{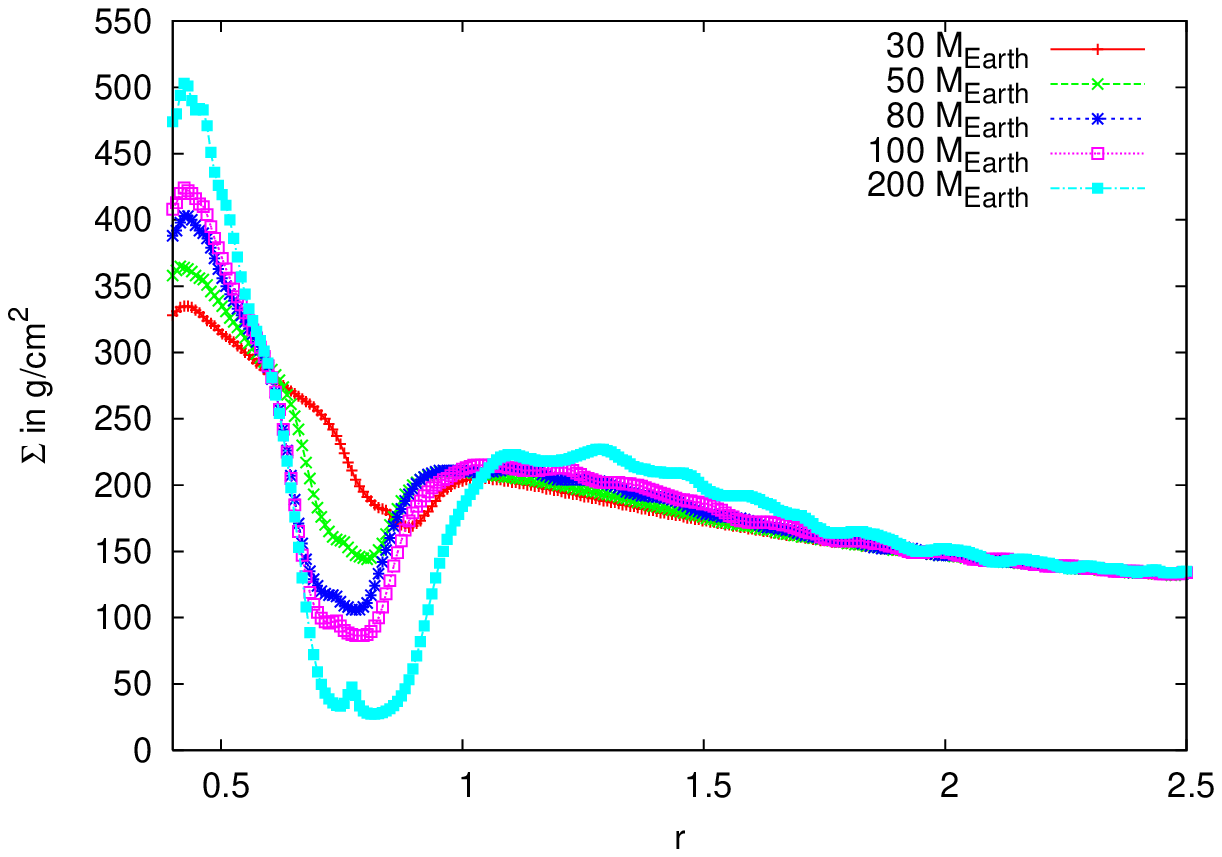}
 \includegraphics[width=0.90\linwx]{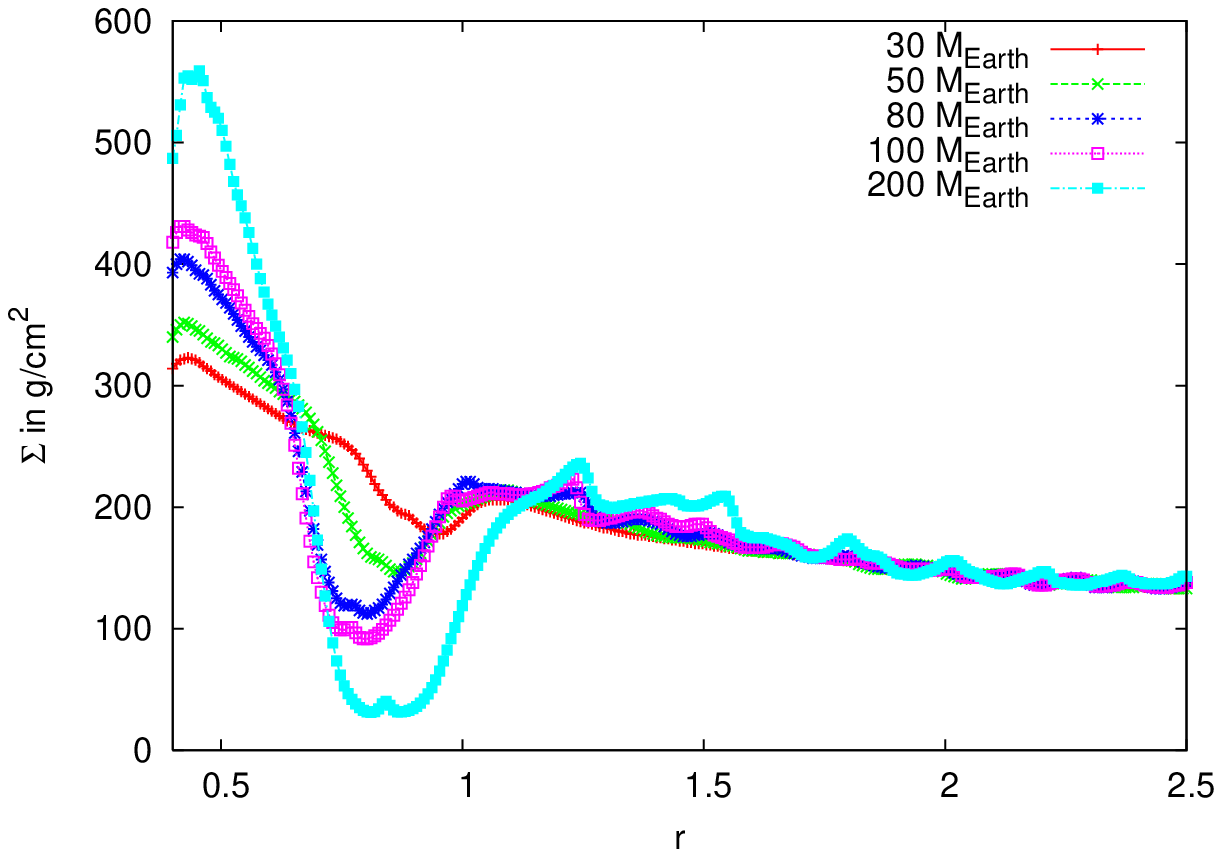}
 \caption{Azimuthally averaged surface density after 200 orbits for isothermal models with different
 planet masses. Top: initial $e_0=0.1$, and Bottom: initial $e_0=0.4$. 
   \label{fig:SigmassIsoe}
   }
\end{figure}

The initial damping of eccentricity for planets with a high initial eccentricity ($e_0=0.40$) depends on the planetary mass, meaning the eccentricity for planets with higher mass is damped faster. This faster eccentricity damping is accompanied by a faster inward migration for higher mass planets. This trends seems to stop as soon as the planets have cleared their gap and the eccentricity damping and evolution of the semi-major axis is nearly the same for all the different high mass planets.
Fig.~\ref{fig:SigmassIsoe} shows the azimuthally averaged surface density at the time of 200 planetary orbits for the two different
initial starting eccentricities. The profiles after 200 orbits look very similar, the largest difference occurs for the 
$200 M_{Earth}$ model which displays clearly a wider gap and slower migration in the long run. For the models with 80 and 100
Earth masses it appears that the migration is slightly faster for the high eccentric case. However, the averaged profile
do not give a clear hint toward the cause.

\subsection{Fully radiative discs}

In the fully radiative regime we can use our suggested cubic potential without problems for higher mass planets as the inclusion of radiation transport/cooling in a disc prevents a large density build-up near the planet, as the temperature near the planet rises and stops further mass accumulation.

\begin{figure}
 \centering
 \includegraphics[width=0.9\linwx]{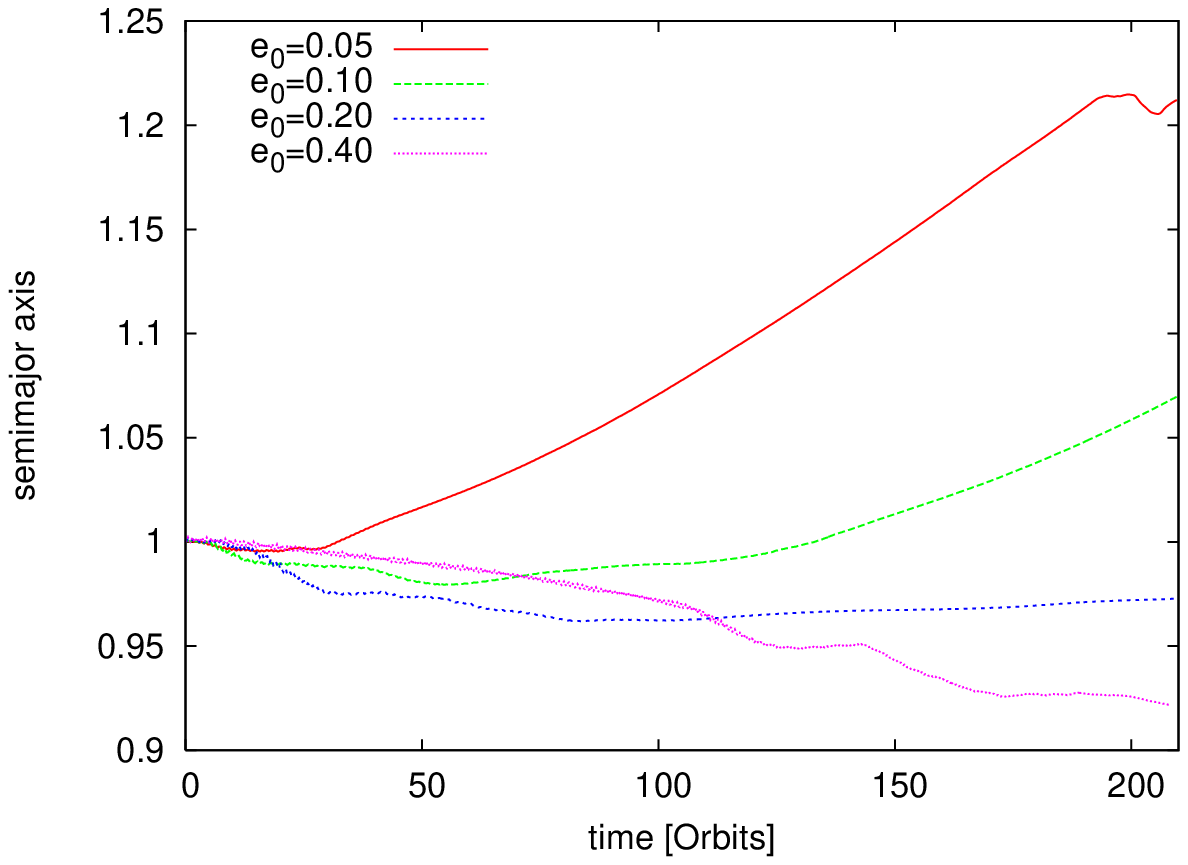}
 \includegraphics[width=0.9\linwx]{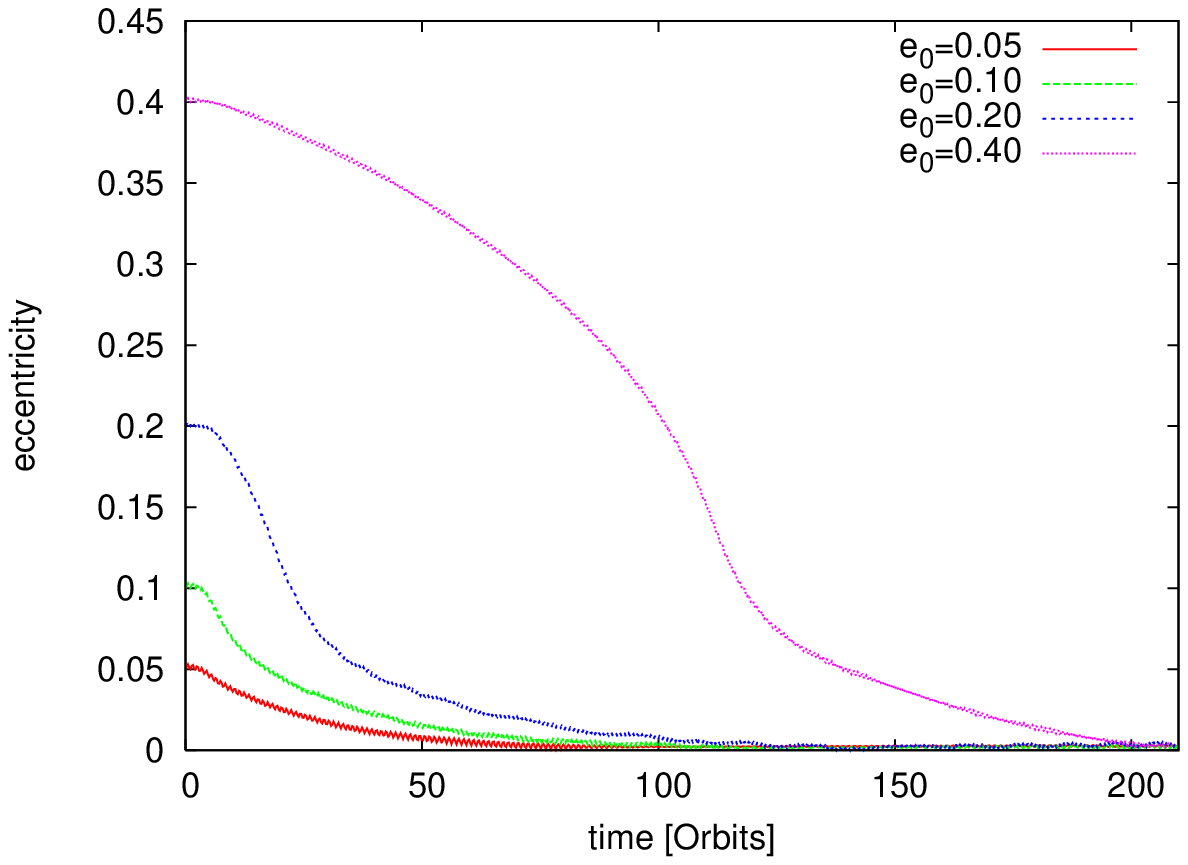}
 \caption{Time evolution of the semi-major axis (top) and eccentricity (bottom) of a $30 M_{Earth}$ planet in a fully radiative disc. The eccentricity shrinks in the same fashion as for a $20 M_{Earth}$ planet, but about $50\%$ faster. The semi-major axis increases also in the same trend, but the outward migration starts about $50\%$ later than in the $20 M_{Earth}$ case. The outward migration is then stopped at a radius slightly smaller than for a $20 M_{Earth}$ planet.
   \label{fig:AExcfullMass30}
   }
\end{figure}

In Fig.~\ref{fig:AExcfullMass30} we display the change of the semi-major axis and eccentricity over time of a $30 M_{Earth}$ planet
for different initial eccentricities. One can clearly see that the outward migration starts when the eccentricity is damped to a small value as we already expected from our results of the $20 M_{Earth}$ planet. The damping of eccentricity is about $50\%$ faster than for the $20 M_{Earth}$ planet. This speed up in the damping of eccentricity is due to the increase of the planet's mass. We also observe a change in the damping rate of the eccentricity as soon as the planet reaches $e \approx 0.10$, as for our previous simulations. For an initial low eccentricity ($e_0 = 0.05$) we observe an earlier outward migration until the planets semi-major axis reaches the aforementioned barrier of outward migration in our discs. This barrier is dependent on the planets mass, as the final semi-major axis is slightly smaller for the $30 M_{Earth}$ planet compared to the $20 M_{Earth}$ planet (see App.~\ref{app:Numfeature}).

\begin{figure}
 \centering
 \includegraphics[width=0.9\linwx]{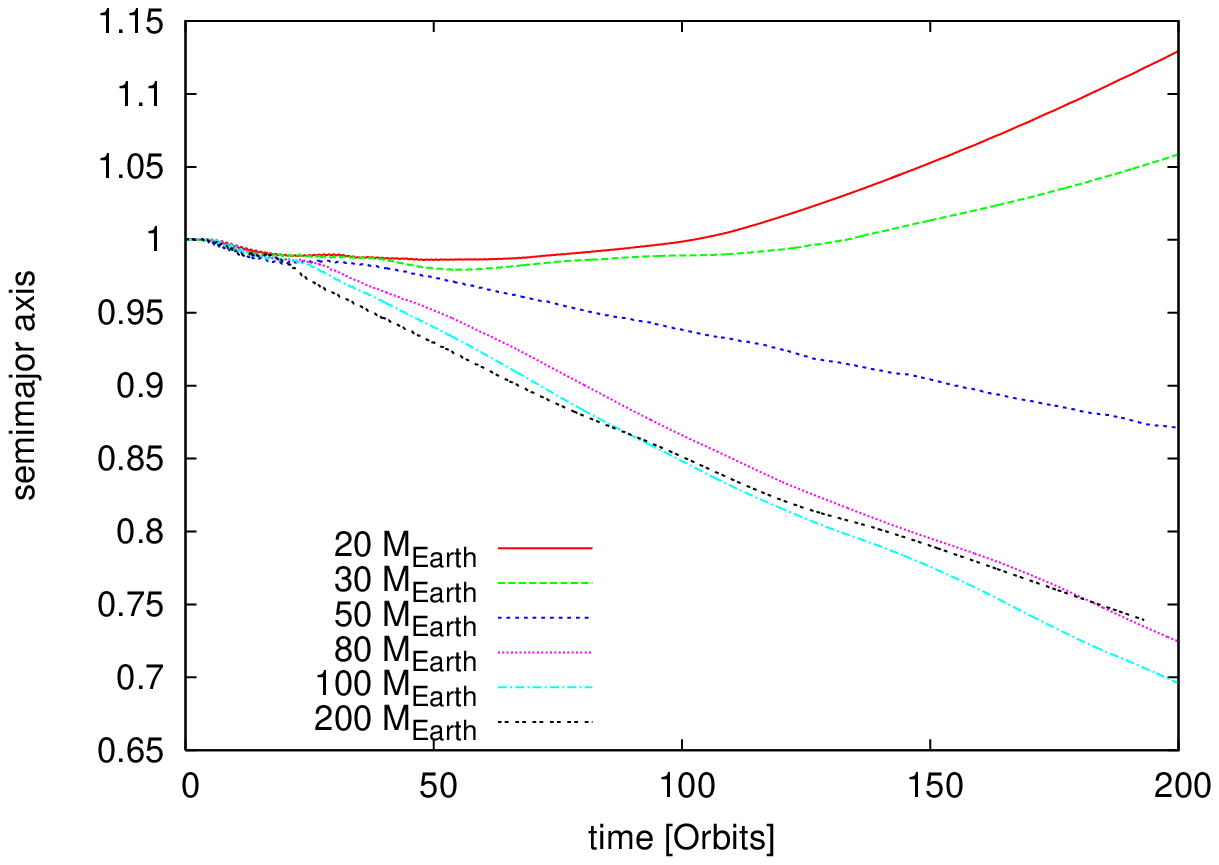}
 \includegraphics[width=0.9\linwx]{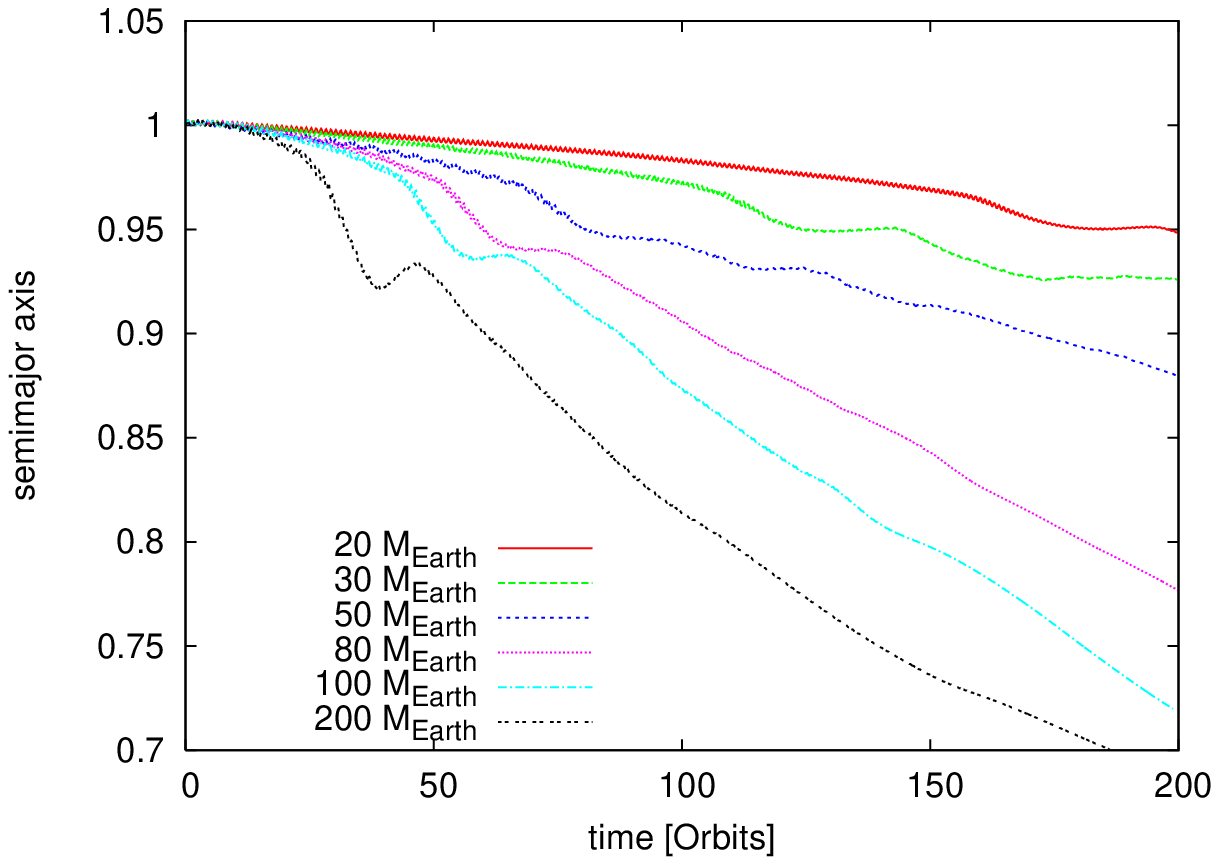}
 \caption{Time evolution of the semi-major axis for planets in a fully radiative disc with $20$, $30$, $50$, $80$, $100$ and $200$ Earth masses. In the top figure the planets have an initial eccentricity of $e_0=0.10$ and in the bottom figure the initial eccentricity is $e_0=0.40$.
   \label{fig:AfullMassAll}
   }
\end{figure}

\begin{figure}
 \centering
 \includegraphics[width=0.9\linwx]{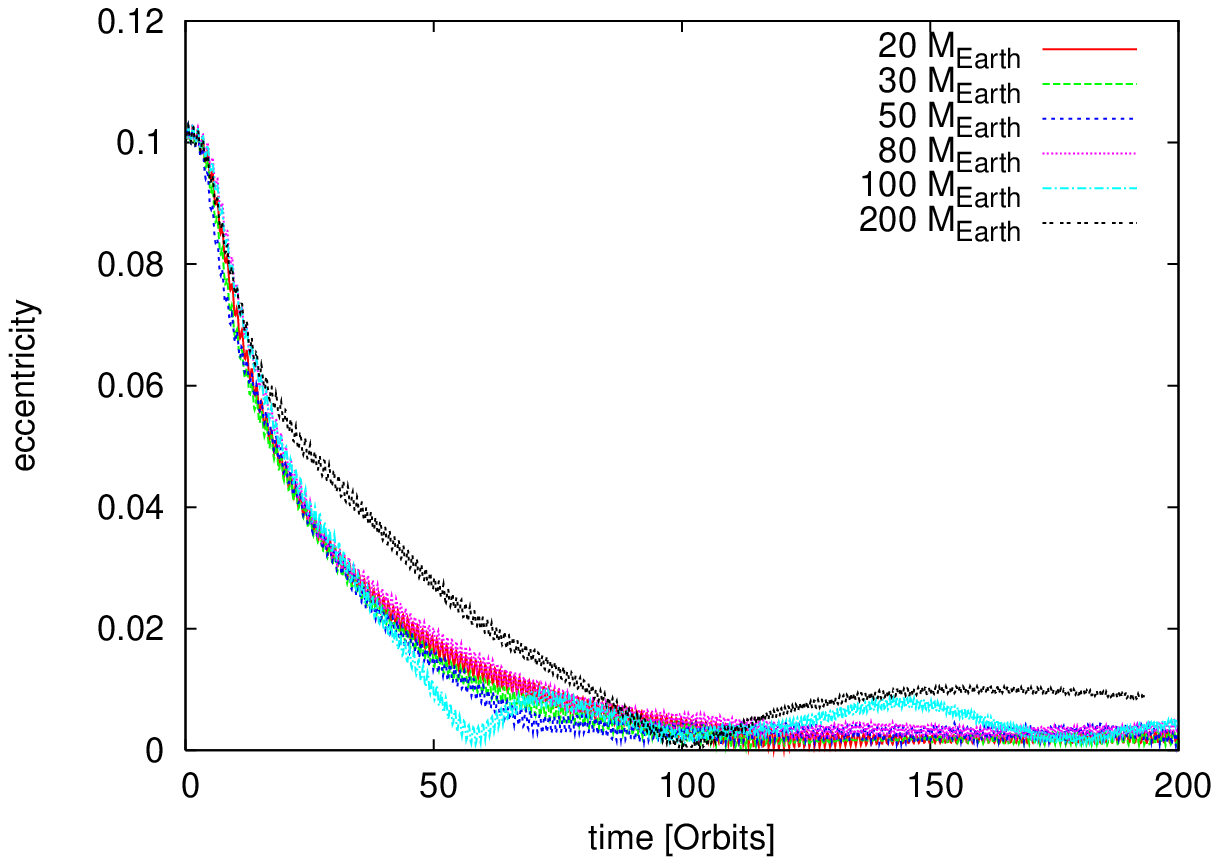}
 \includegraphics[width=0.9\linwx]{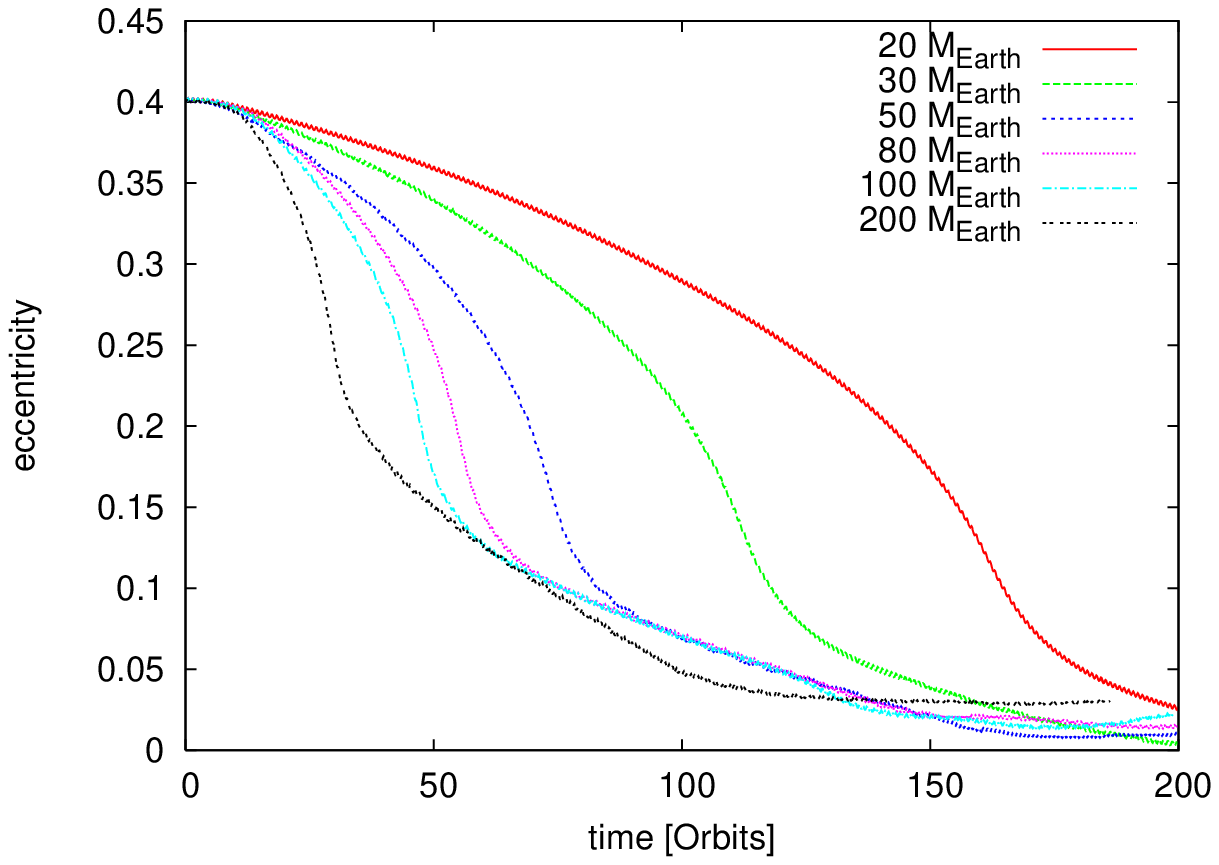}
 \caption{Time evolution of the eccentricity for planets in a fully radiative with $20$, $30$, $50$, $80$, $100$ and $200$ Earth masses. In the top figure the planets have an initial eccentricity of $e_0=0.10$ and in the bottom figure the initial eccentricity is $e_0=0.40$.
   \label{fig:EccfullMassAll}
   }
\end{figure}

In our previous work we obtained a limiting planet mass of about $33 M_{Earth}$ for the occurrence of outward migration \citep[see][]{2009A&A...506..971K}.
This implies that planets with a higher mass will not migrate outward but inward in a fully radiative disc. In Fig.~\ref{fig:AfullMassAll} we display the evolution of the semi-major axis of planets with $20$, $30$, $50$, $80$, $100$ and $200$ Earth masses for planets with an initial eccentricity of $e_0=0.10$ and $e_0=0.40$. In Fig.~\ref{fig:EccfullMassAll} we display the eccentricity for the same set of parameters.

For the planets with $20$ and $30$ Earth masses we observe outward migration as soon as the eccentricity is damped to a small value ($e \leq 0.02$). As this damping is only achieved for the $e_0=0.1$ simulations during the run-time of our simulations, we can not see outward migration in the $e_0=0.40$ figure, but we have seen it for the $20 M_{Earth}$ planet in Fig.~\ref{fig:AfullEccall}. 

The planets with $50$, $80$, $100$ and $200$ Earth masses migrate always inward, independent of the initial eccentricity value. However the inward migration is much slower for the $50 M_{Earth}$ planet compared to the high mass counterparts. The $80$, $100$ and $200 M_{Earth}$ planets migrate inward with the fastest rate, but the relative speed of inward migration for these three planets does not differ that much as it does for the $50$ and $80 M_{Earth}$. If high mass planets ($M > 50 M_{Earth}$) have an initial higher eccentricity they migrate inward a little bit slower than their low eccentric counterparts. As an eccentric orbit disrupts the typical spiral wave structure of a circular orbit the migration rate of a planet on an eccentric orbit is altered compared to the migration rate of a planet on a nearly circular orbit. Having reached a nearly circular orbit the effects of radiation transport/cooling set in and the planet can (if its mass is low enough) migrate outward. 

In Fig.~\ref{fig:EccfullMassAll} we display the eccentricity evolution for planets starting with $e_0=0.1$ and $e_0=0.4$. The eccentricity damping for planets starting with $e_0=0.10$ seems to be independent of the planet's mass, if $M_P \geq 25 M_{Earth}$. The damping sets in immediately after the planets reach their final mass and the eccentricity is damped to nearly the same value for all simulations. The planet's mass seems to have little effect on eccentricity damping for planets with a low initial eccentricity. The $100$ and $200 M_{Earth}$ planet show little fluctuations in the eccentricity when the eccentricity reaches zero. These fluctuations have their origin in the fact that this planet lies very close to the inner boundary of our simulation (we use reflective boundary conditions at the inner boundary) and thus interacts with it which give rise to the fluctuations in the eccentricity. These fluctuations are also enhanced by the planet's mass.

On the other hand, if the planets have an initial high eccentricity (e.g. $e_0=0.40$), we observe a quite different damping rate of the eccentricity. The damping is faster for planets with higher mass, but the final eccentricity reached is the same for all planetary masses in our simulations. As soon as the $100$ and $200 M_{Earth}$ planet reach an eccentricity of $e\approx 0.3$ they loose half their eccentricity in a time of a few orbits, which also affects the migration rate of the planet, as we can observe a little bump at the same time. As in the isothermal case the initial damping rate of eccentricity is reduced for smaller mass planets. Interestingly, all planets experience a similar $\dot{e}$-rate once the eccentricity has dropped below about $0.1-0.15$.

As for planets with $20 M_{Earth}$ an initial eccentricity influences the migration of planets with higher masses. In case the planet is small enough to undergo outward migration the effect of an initial eccentricity is a halting of outward migration to the point the eccentricity is damped to a very small value so that the effects of radiation transport/cooling can take effect as if the planet was moving on a circular orbit. For planets with a mass so high that they will not undergo outward migration an initial eccentricity can slow down the inward migration process by only a very small amount.

\subsection{The criterion for outward migration}

In the previous simulations we have seen that planets can migrate outward only for sufficiently small orbital eccentricities.
The occurrence of outward migration is linked to the detailed structure of the horse-shoe
region since the torques originate from a region with a very small radial extent \citep{2009A&A...506..971K}. It is to be expected that
eccentric orbits will destroy the detailed balance, and this is what we see in our simulations. Nevertheless, it is interesting to estimate
the value of eccentricity at which this reversal of migration can occur. For that purpose we have run additional series of test simulations
in only two spatial dimensions but with identical physical disc parameter for various planetary masses. We measured the limiting value of the eccentricity with two alternative methods. In the first set we performed simulations with planets on fixed orbits for different eccentricities and
masses. The point of migration reversal (equivalent to a sign-reversal of the power) has then been determined from the time averaged torque and power
measured after $100$ orbits. The time averaging has been done over 5 orbits. In the second alternative we followed the orbital evolution of the planet
in the disc starting with an initial eccentricity of $e_0=0.10$. As demonstrated above all planets reduce this initial eccentricity and will migrate
outward in a radiative disc after the eccentricity has dropped below a certain value. From the time evolution we determine this critical
eccentricity. This second method is used for our full 3D disc as well.

\begin{figure}
 \centering
 \includegraphics[width=0.9\linwx]{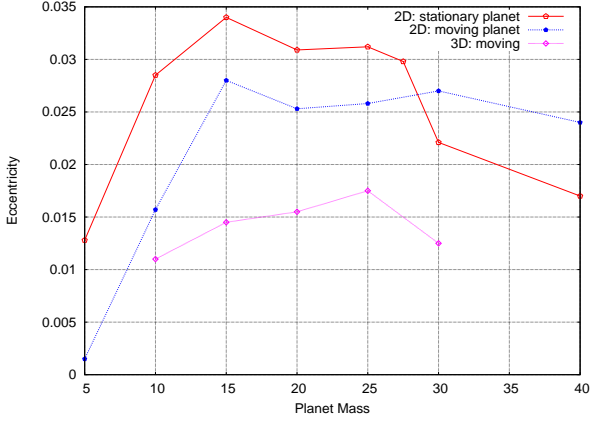}
 \caption{The critical eccentricity for reversal of migration obtained using 2D and 3D simulations for fully radiative discs.
 Only planets having eccentricities below this curve are prone to outward migration.
   \label{fig:ecc-crit}
   }
\end{figure}

The results obtained using these procedures are displayed in Fig.~\ref{fig:ecc-crit} for the three sets of simulations. While
the general trend is similar in all three series there are nevertheless differences. The 2D results have an average value of
$e_{crit} = 0.027$ with a drop for smaller mass planets. The values are systematically larger than those of the 3D runs with $e_{crit} = 0.015$.
This effect may be caused by the slightly higher temperature in the disc ([$H/r]_{2D} \approx 0.045$) compared to the
3D runs or due to genuine flow differences in 2D and 3D geometry.
Due to this larger $H/r$ planets may experience outward migration for higher masses as well in a 2D geometry \citep{2008A&A...487L...9K}.
Within the 2D runs the results obtained for fixed stationary orbits yield slightly larger values for smaller masses
and drop below the results for evolving planets for masses above $25 M_{Earth}$. In \citet{2009A&A...506..971K} we have observed that the location of the
torque maximum responsible for the outward migration lies slightly offset of the planet location at $r\approx 0.984$.
Interestingly, for an eccentricity of $e=0.015$ the apoastron lies close to this position, and we expect a destruction of this effect.
So the planet can only migrate outward if its eccentricity is smaller than the distance to the offset in the torque distribution,
which explains why the fully radiative effects do not turn the planet to outward migration if it has a high eccentricity.
This eccentricity has to be damped below this value before the planet can undergo outward migration.
The offset distance appears to be relatively insensitive to the mass of the planet \citep{2009A&A...506..971K} and does
not scale directly with the planet's Hill radius. \citet{2008ApJ...672.1054B} even suggest a torque maximum directly at the location
of the planet. Our simulations seem to indicate a dependence on the thermodynamic structure of the disc, such as radiative diffusion and temperature
(entropy) gradients. This issue still requires resolution. 

\section{Summary and Conclusions}

We have performed full 3D radiation hydrodynamical simulations of accretions discs with embedded planets of different masses on eccentric orbits.
In a first sequence of simulations we have analysed in detail the dynamics of a planet with a given mass of $20 M_{Earth}$ for the isothermal as well as fully radiative case. In the isothermal situation we studied the cases $H/r=0.05$ (a value often used in planet-disc simulations) and $H/r=0.037$
(the value that matches the fully radiative case). In both cases we find similar behaviour for the eccentricity and semi-major axis evolution,
and results that match those of \citet{2007A&A...473..329C} very well.
Small eccentricities (with $e \lsim 2 H/r$) are damped exponentially with a time scale given approximately by the linear results \citep{1994Icar..110...95W}.
Larger eccentricities are damped initially according to $\dot{e} \propto e^{-2}$ in agreement with \citet{2000MNRAS.315..823P} and \citet{2007A&A...473..329C}.
The final value of the eccentricity does not depend on the initial eccentricity of the planet.

The planet migrates inward in the isothermal regime.
Low mass planets (e.g. $20 M_{Earth}$) on eccentric orbits with large eccentricity ($e>0.20$) have a slower migration rate as their low eccentric counterparts in the isothermal regime. As soon as the damping of eccentricity proceeds to smaller values the migration rate is pumped up as if the planet had started with a low eccentricity. The maximum inward migration rate occurs at $e \approx 2 H/r$.
In the fully radiative regime high eccentric planets ($e>0.20$) with $20 M_{Earth}$ migrate inward on a rate comparable with the isothermal regime. The corresponding eccentricity damping rate for the fully radiative scheme is about the same as for the isothermal simulations, taking into account the different sound speeds.

But as soon as the eccentricity becomes small enough the planets undergo a change in the direction of migration.  
The inclusion of radiation transport/cooling in discs with embedded low mass planets will give rise to a change in the direction of migration for planets whose initial eccentricity has been damped to a nearly circular orbit. The maximum eccentricity a planet can have to still undergo outward migration seems to be determined by the torque maximum in our $\Gamma (r)$ function. This torque maximum has a slight offset compared to the planets location as demonstrated in \citet{2009A&A...506..971K}, corresponding to a limiting eccentricity of about 0.015-0.025. 
If the eccentricity of the planet is larger than this value it will migrate inward, while it will migrate outward if its eccentricity is smaller (see Fig.~\ref{fig:ecc-crit}). For very small planet masses the maximum eccentricity is reduced. For planets on nearly circular orbits the effects of radiation keep the positive torques acting on the planet unsaturated, which implies continuous outward migration. Moving planets experience this result directly, and do indeed migrate outward in the disc in contrast to planets in the isothermal regime. 

Eccentric planets with higher mass are slowed down in their migration at the beginning if they have a high initial eccentricity ($e_0 \geq 0.20$) in the isothermal as well as in the fully radiative scheme. If $e \geq 0.02$ all planets move inward independent of their mass, even those embedded in a fully radiative disc. When the eccentricity is damped further the high mass planets ($M \geq 50 M_{Earth}$) still move inward for both regimes, as they open a gap in the disc and migrate as Type II. The eccentricity damped $30 M_{Earth}$ planet moves outward in the fully radiative scheme as we expected from our previous results \citep{2009A&A...506..971K}. 

Independent of the discs thermodynamics and planet mass, we find that an embedded planet with a given initial eccentricity will lose this eccentricity in time. The rate of the eccentricity loss depends on the value of the initial eccentricity but is much faster than the migration time. 
Hence, according to our results planet-disc interaction cannot be the cause of the observed high mean eccentricity of extrasolar planets. 
This finding is supported by the fact that the existence of planetary systems in mean-motion resonance with small libration angles require
damping of eccentricity \citep{2002ApJ...567..596L, 2005A&A...437..727K}.
A solution to this problem might be planet-planet interaction, which we have not considered here. 

While performing our studies we noticed that numerical resolution is a serious issue in these type of simulations. As shown in the appendix, only for very
high numerical resolution or in an adapted rotating coordinate system which rotates with the present location of the planet, can we observe the
outward migration.
Finally, as the origin of this outward migration for planets below $M_p \approx 33 M_{Earth}$ is created by a delicate balance of torques which
is destroyed by even a very small eccentricity of the planet, the question arises how this effect can persist under realistic conditions.
It remains to be studied what effect the turbulent motions within the disc have on the corotation torque of the planet.

\begin{acknowledgements}

B. Bitsch has been sponsored through the German D-grid initiative. W. Kley acknowledges the support through the German Research Foundation (DFG) through grant KL 650/11 within the Collaborative Research Group FOR 759: {\it The formation of Planets: The Critical First Growth Phase}. The calculations were performed on systems of the Computer centre of the University of T\"ubingen (ZDV) and systems  operated by the ZDV on behalf of bwGRiD, the grid of the Baden  W\"urttemberg state. 

\end{acknowledgements}

\appendix
\section{Numerical features for moving Planets}
\label{app:Numfeature}

\begin{figure}
 \centering
 \includegraphics[width=0.9\linwx]{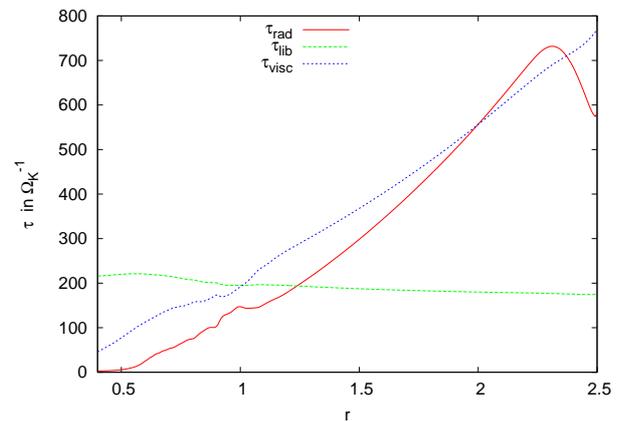}
 \caption{Timescales in dependence of the distance from the central star. To compute the timescales we used the density and temperature of the midplane at the beginning of the fully radiative simulations.
   \label{fig:tauall}
   }
\end{figure}

In Fig.~\ref{fig:AfullEccall} we have noticed that the outward migration of the moving 20 $M_{Earth}$ planet terminates at a well defined radius
independent of its initial eccentricity.
To determine theoretically the extent of outward migration from our disc properties it seems useful to compare different timescales: 
the libration time and the radiation time scale of the disc.
The necessary unsaturated torques needed for sustained outward migration require approximately equal libration and cooling
time (see \citet{2008ApJ...672.1054B}; \citet{2008arXiv0804.4547P}). 
For the latter we use in our case the radiative diffusion time scale. We define
\begin{equation}
  	\tau_{rad}=\frac{s^2}{D_{rad}}
\end{equation} 
with $s=H$ and
\begin{equation}
	D_{rad}  =  \frac{4 c a T^3}{3 c_v \rho^2 \kappa} \ .
\end{equation}
The libration time is (as in \citet{2008ApJ...672.1054B}) $\tau_{lib} = 8 \pi r_{P} / (3 \Omega_K x_s)$ with the half-width of the horseshoe-orbit $x_s=1.16 r_P \sqrt{q / (H/r)}$. To compute the timescales we use the density and temperature at the midplane of the disc at the beginning of the simulations. The timescales are displayed in Fig.~\ref{fig:tauall}. 
The timescales seem to be comparable to about $r=1.4$, so the planet should be able to migrate outward at least to this point and should not stop at $r=1.23$ as observed in Fig.~\ref{fig:AfullEccall}. 
In the plot we also show additionally the viscous timescale $\tau_{visc} = s^2/\nu$ which should be comparable to the
radiative time for accretion discs in equilibrium. Apparently, this relation is well fulfilled.

The migration of the planet inside the disc is calculated via the torque acting on the planet. In our previous work \citep{2009A&A...506..971K} we found that the calculated torque acting on the planet appears to be converged already for our standard resolution if the simulation is performed in a rotating coordinate system where the planet is fixed for circular orbits. If the planet is allowed to move freely inside the disc,
for example due to the disc's gravitational force, this is no longer possible, as the planet changes its semi-major axis during time. If the planet moves to an orbit with larger radius the rotating frame will not be able to keep the planet at the same point in the grid, as the rotation frequency of the frame is not linked to the planet anymore.

\begin{figure}
 \centering
 \includegraphics[width=0.9\linwx]{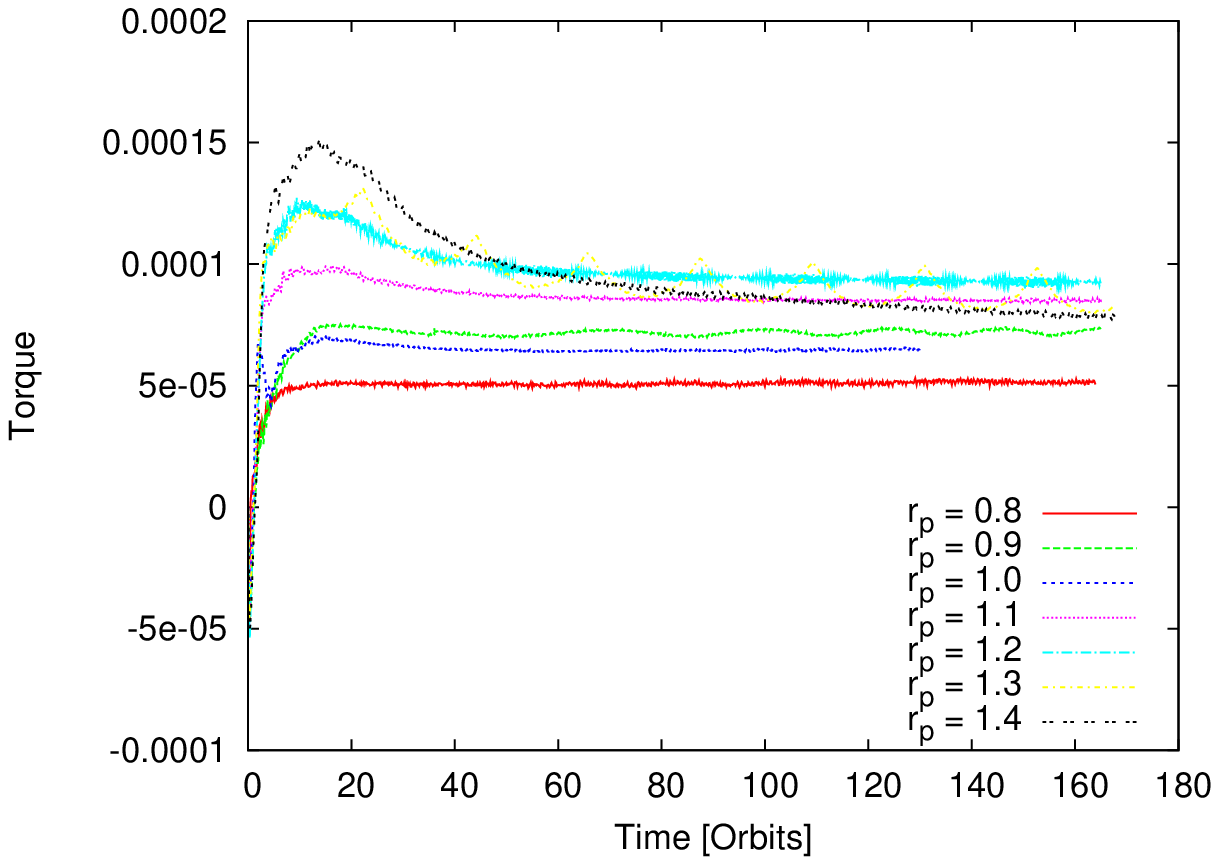}
 \includegraphics[width=0.9\linwx]{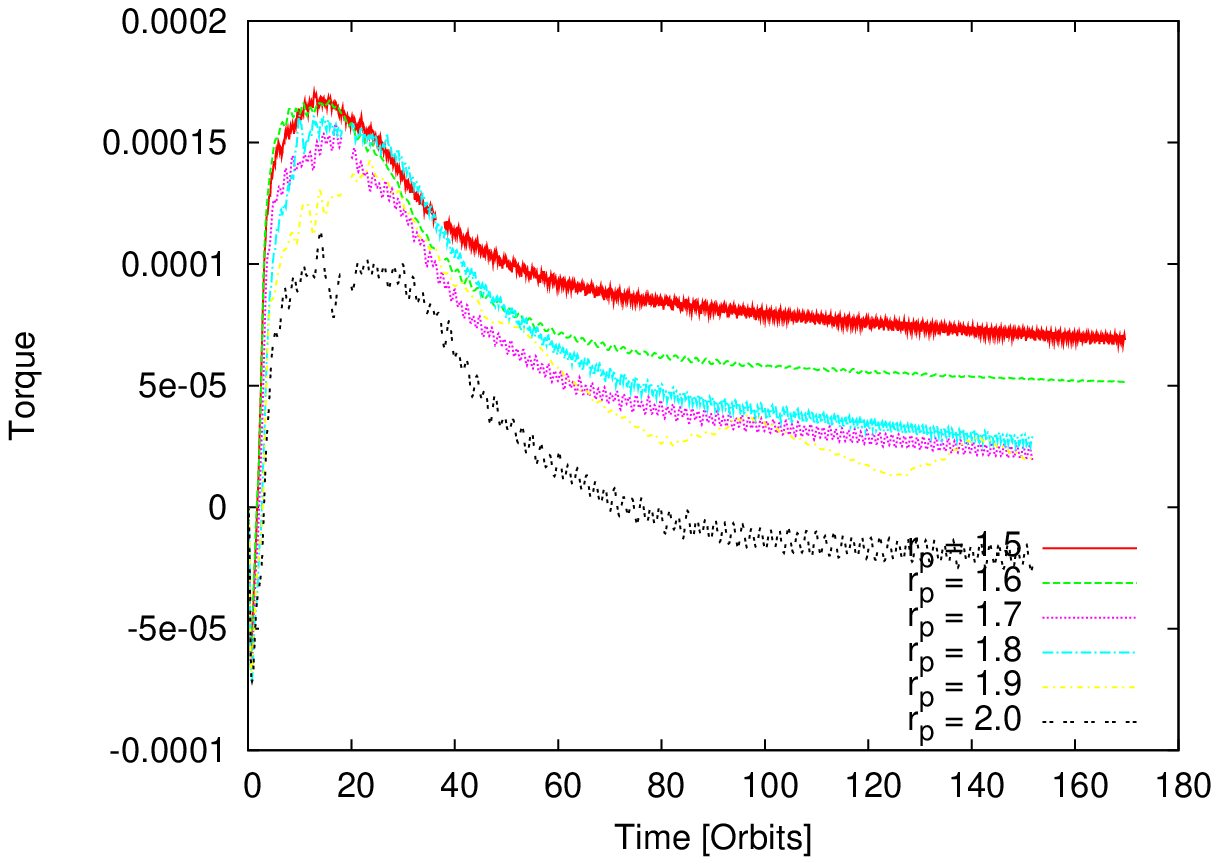}
 \caption{Torque for planets on fixed circular orbits with distances $r_p = 0.8 r_{Jup}$ to $r_p = 2.0 r_{Jup}$.
 In each simulation the coordinate system is rotated with the planet.
   \label{fig:Torque08bis20}
   }
\end{figure}

To investigate this, we performed a series of simulations with a  $20 M_{Earth}$ planet on a fixed circular orbits placed at
various distances from the star. In each of the runs the coordinate system rotated with the orbital period of the planet
such that the planet did not move through the grid. 
In Fig.~\ref{fig:Torque08bis20} we display the torques acting on this planet in the fully radiative scheme. 
The planets are placed in a disc, corresponding to our standard model, at different planetary radii ranging from
$r_p = 0.8 r_{Jup}$ to $r_p = 2.0 r_{Jup}$. 
As the rotation frequency of the planet matches the rotation frequency of the coordinate system, the planet remains at a
fixed location in the grid during the whole simulations. Clearly, the torque acting on these planets is positive for planets 
with $r_p$ up to $r_p = 1.9 r_{Jup}$, indicating that a planet at these radii should still migrate outward. 
The positive torques are contrary to the results of the moving planets in Fig.~\ref{fig:AfullEccall}, where the planets
stop their outward migration already at about $r=1.23$.

\begin{figure}
 \centering
 \includegraphics[width=0.9\linwx]{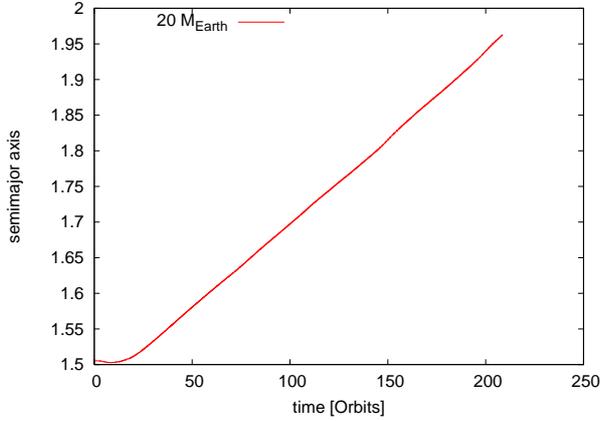}
 \caption{Semi-major axis of a $20 M_{Earth}$ planet starting at $r_p=1.5$ with a grid rotating with the initial angular speed of
 the planet.
   \label{fig:r15w054}
   }
\end{figure}

The torques acting on theses planets indicate, that a planet in the disc should migrate outward until at least $r=1.9$.
Or possibly even further, if not stopped by our finite computational domain where numerical effects from the outer boundary may disturb its way.
To test this hypothesis we calculated the evolution of a $20 M_{Earth}$ planet starting at $r=1.5$ with a rotation frequency
of the grid matching the rotation frequency of the planet at this distance ($r=1.5$).
Indeed, as Fig.~\ref{fig:r15w054} shows, the planet now
migrates outward as the torques presented in Fig.~\ref{fig:Torque08bis20} suggested. Note that the mass of the planet
has been increased gradually within the first 10 orbits.
Again, this result is in contradiction to the stopped outward migration in Fig.~\ref{fig:AfullEccall}.

\begin{figure}
 \centering
 \includegraphics[width=0.9\linwx]{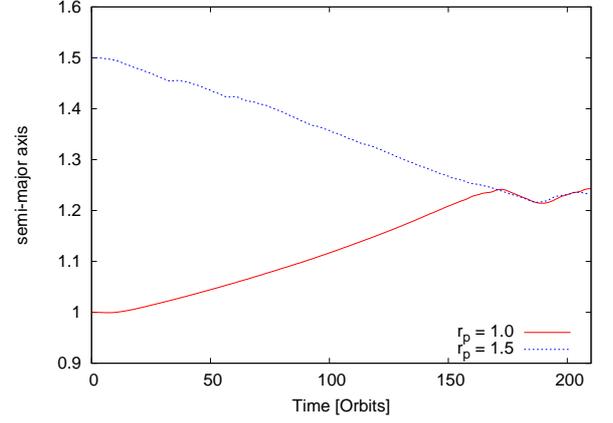}
 \caption{Semi-major axis of planets starting at $r=1.0$ and $r=1.5$ with a rotation frequency of the numerical grid matching the Keplerian rotation
 at $r=1.0$ for both cases.
   \label{fig:r123}
   }
\end{figure}

May this effect be caused by the difference in the rotation frequency of the planet and the numerical grid? 
To answer this, we first place a $20 M_{Earth}$ planet at $r_p=1.5$ with a rotation frequency corresponding to a $r=1.0$ and let it evolve
in the disc. The evolution of the semi-major axis is displayed in Fig.~\ref{fig:r123}.
Secondly, we started an identical planet at $r_p=1.0$ with the same rotation frequency of the numerical grid. 
In the end of the evolution both planets comes to a halt at the same radius. 
Obviously, the chosen rotation frequency of the numerical grid has an influence of the migration of the planet inside the disc.
The reason lies in the increased numerical diffusion that occurs if matter moves with a fast speed through the disc. The implemented
{\tt FARGO} algorithm helps to solve this problem but cannot fully eliminate it. The problem is enhanced in our situation because it is the
detailed fine structure in the flow near the planet that determines the outcome of migration.

\begin{figure}
 \centering
 \includegraphics[width=0.9\linwx]{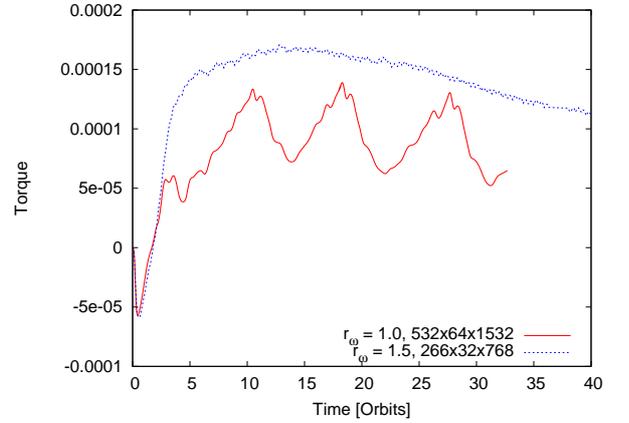}
 \caption{Torque acting on a $20 M_{Earth}$ planet at $r_p =1.5$. Red solid line: The rotation frequency of the grid matching the Keplerian rotation
  frequency at $r=1.0$, and the grid resolution has been doubled compared to our standard simulations. Blue dotted line: Standard resolution with
  grid velocity equal to the planet.
   \label{fig:r15Double}
   }
\end{figure}

Being caused by numerical diffusion, the effect may be (partially) cured by increasing the resolution of the grid. We now double the grid size to $532 \times 64 \times 1532$ active cells in $r, \theta, \phi$ direction and let a planet fixed at $r_p =1.5$ evolve with a rotation frequency corresponding to $r=1.0$.
The torque acting on the planet is displayed in Fig.~\ref{fig:r15Double} and is clearly positive, indicating outward migration. Over-plotted is the torque of a planet at $r_p=1.5$ with matching rotation frequency and our standard resolution (the same as in Fig.~\ref{fig:Torque08bis20}).

Hence, it seems that the rotation frequency of the numerical grid influences the migration of the planet in the disc, if the numerical resolution of the grid is too small and the planet moved to a radius where its rotation frequency differs by more than $25 \%$ from rotation frequency of the grid. In that way, our results determined in the main article are correct, as the planets have migrated inside the disc only by a little bit before their eccentricity is damped and they start their outward migration. The obtained limit of $r=1.23$ for the outward migration seems to be a numerical artefact, however.
To obtain an accurate migration for planets under these conditions it seems best to perform the simulations in a coordinate system that rotates always with the speed of the planet. For multiple
planetary systems this is not possible and a higher resolution is required.

\bibliographystyle{aa}
\bibliography{kley8}
\end{document}